\documentclass[%
 reprint,
 amsmath,amssymb,
 aps,
]{revtex4-1}
\usepackage{graphicx} 

\begin{document}


\title{Variations in the Fundamental constants in the QSO Host J1148+5251 at $z = 6.4$ and the BR1202$-$0725 System at $z = 4.7$.}


\author{Lindley Lentati}
\email[]{ltl21@cam.ac.uk}
\affiliation{Astrophysics Group, Cavendish Laboratory, JJ Thomson Avenue,  Cambridge, CB3 0HE, UK}
\author{C. Carilli}
\affiliation{National Radio Astronomy Observatory, P.O. Box O, Socorro, NM 87801, USA}
\author{P. Alexander}
\affiliation{Astrophysics Group, Cavendish Laboratory, JJ Thomson Avenue,  Cambridge, CB3 0HE, UK}
\author{R. Maiolino}
\affiliation{Astrophysics Group, Cavendish Laboratory, JJ Thomson Avenue,  Cambridge, CB3 0HE, UK}
\author{R. Wang}
\affiliation{National Radio Astronomy Observatory, P.O. Box O, Socorro, NM 87801, USA}
\author{P. Cox}
\affiliation{Institut de Radio Astronomie Millimetrique, 300 Rue de la Piscine, Domaine Universitaire, 38406 Saint Martin d'H\'{e}res, France}
\author{D. Downs}
\affiliation{Institut de Radio Astronomie Millimetrique, 300 Rue de la Piscine, Domaine Universitaire, 38406 Saint Martin d'H\'{e}res, France}
\author{R. Neri}
\affiliation{Institut de Radio Astronomie Millimetrique, 300 Rue de la Piscine, Domaine Universitaire, 38406 Saint Martin d'H\'{e}res, France}
\author{R. McMahon}
\affiliation{Institute of Astronomy, University of Cambridge, Madingley Road, Cambridge, CB3 0HA, UK}
\author{K. M. Menten}
\affiliation{Max-Planck-Institut f\"{u}r Radioastronomie, Auf dem H\"{u}gel 69, 53121 Bonn, Germany}
\author{D. Riechers}
\affiliation{Astronomy Department, California Institute of Technology, MC 249-17, 1200 East California Boulevard, Pasadena, CA 91125, USA}
\author{J. Wagg}
\affiliation{European Southern Observatory, Alonso de Cordova 3107, Vitacura, Casilla 19001, Santiago 19, Chile}
\altaffiliation{Astrophysics Group, Cavendish Laboratory, JJ Thomson Avenue,  Cambridge, CB3 0HE, UK}
\author{F. Walter}
\affiliation{Max-Planck-Institut f\"{u}r Astronomie, K\"{o}nigstuhl 17, D-69117 Heidelberg, Germany}
\author{A. Wolfe}
\affiliation{Department of Physics, Centre for Astrophysics \& Space Sciences, University of California, San Diego, CA 92093 USA}


\date{\today}
\begin{abstract}
We use sensitive observations of three high redshift sources; [C~II] $^2P_{3/2} \to ^2P_{1/2}$ fine structure and CO~(J = $2\to1$) rotational transitions for the $z=6.4$ Quasar host galaxy (QSO) J1148+5251 taken with the Plateau de Bure Interferometer (PdBI) and Jansky Very Large Array (JVLA) respectively, and [C~II] and CO~(J = $5\to4$) transitions from the QSO BR1202$-$0725 and its sub-millimeter companion (SMG) galaxy at $z=4.7$ taken with the Atacama Large Millimetre Array (ALMA) and the PdBI.  We use these observations to place constraints on the quantity $\Delta z = z_{CO} - z_{C_{II}}$ for each source where $z_{CO}$ and $z_{C{II}}$ are the observed redshifts of the CO rotational transition and [C~II] fine structure transition respectively, using a combination of approaches; 1) By modelling the emission line profiles using `shapelets' - a complete ortho-normal set of basis functions that allow us to recreate most physical line shapes - to compare both the emission redshifts and the line profiles themselves, in order to make inferences about the intrinsic velocity differences between the molecular and atomic gas, and 2) By performing a marginalisation over all model parameters in order to calculate a non-parametric estimate of $\Delta z$.  We derive 99$\%$ confidence intervals for the marginalised posterior of $\Delta z$ of $(-1.9 \pm 1.3) \times 10^{-3}$, $(-3 \pm 8) \times 10^{-4}$ and $(-2 \pm 4) \times 10^{-3}$ for J1148+5251, and the BR1202$-$0725 QSO and SMG respectively.   We show the [C~II] and CO~(J = $2\to1$) line profiles for J1148+5251 are consistent with each other within the limits of the data, whilst the [C~II] and CO~(J = $5\to4$) line profiles from the BR1202$-$0725 QSO and SMG respectively have $65$ and $>99.9\%$ probabilities of being inconsistent, with the CO~(J = $5\to4$) lines $\sim 30\%$ wider than the [C~II] lines.  Therefore whilst the observed values of $\Delta z$ can correspond to variations in the quantity $\Delta F/F$ with cosmic time, where $F=\alpha^2/\mu$, with $\alpha$ the fine structure constant, and $\mu$ the proton-to-electron mass ratio, of both $(-3.3 \pm 2.3) \times 10^{-4}$ for a look back time of 12.9 Gyr and of $(-5 \pm 15) \times 10^{-5}$ for a look back time of 12.4 Gyr we propose that they are the result of the two species of gas being spatially separated as indicated by the inconsistencies in their line profiles.
\newline
\newline
\end{abstract}


\maketitle

\section{Introduction}

One of the most important principles that underlies much of modern physics is that of Copernicus; that our location in space and time is not special, such that the laws of physics do not change from one point in spacetime to another.
All these laws contain within them a relationship to the fundamental constants; electromagnetic interactions involve the fine structure constant $\alpha$, gravitational interactions depends on the gravitational constant $G$ and so on.  It is therefore a natural question to ask whether these constants are constant with cosmic epoch.

This question is not a new one, being proposed some 70 years ago by Dirac \cite{1937Natur.139..323D} and Milne \cite{1935QB500.M5.......}.  Since then however it has become of particular interest with the development of physical models that seek to unify the fundamental forces via extra dimensions, such as Kaluza--Klein, or superstring theories (see e.g.  \cite{2003RvMP...75..403U}, for a review of theoretical models).

These extra dimensions would have only manifested themselves in the first instant of time following the Big Bang ($\sim 10^{-43}$ seconds, or energies of $\sim 10^{19}$ GeV) before compactifying to the Planck scale during the cosmic expansion that followed.

The constants of nature as we observe them in our 4 dimensional spacetime can be shown to depend on their integral properties over all existing dimensions \cite{1990eaun.book.....K}, and as such variations in the scale factors of these now compactified extra dimensions over cosmic time could manifest themselves physically in our 4D Universe as an evolution of the fundamental constants.

Whilst the nature of the potential evolution of these extra dimensions is not well understand, with 
models existing for increasing, decreasing, or oscillating scale lengths (see e.g. \cite{1984PhRvL..52..489M,2004PhLB..581..141M,2005IJMPD..14..677F}), a measurement showing any evolution of the constants would provide supporting evidence for the existence of compact dimensions. 

The current best terrestrial estimates on the evolution of $\alpha$ and $\mu$, the proton to electron mass ratio, come from \citet{2008Sci...319.1808R} of $|\dot{\alpha}/\alpha| = (1.6 \pm 2.3)\times10^{-17}$ yr$^{-1}$ and \citet{2008PhRvL.100n0801B} of  $|\dot{\mu}/\mu| = (1.6 \pm 1.7)\times10^{-15}$ yr$^{-1}$. 

\begin{table*}
\centering
\caption{Existing Constraints on the Evolution of Physical Constants} 
\centering 
\begin{tabular}{c c c} 
\hline\hline 
Physical Quantity ($X$) & $z$ & $\Delta X/X$ \\[0.5ex] 
\hline 
$\alpha$						& 1.84 & $(5.4 \pm 2.5) \times 10^{-6}$ \cite{2007AA...466.1077L} \\
$\alpha$						& 1.15 & $(-0.12 \pm 1.79) \times 10^{-6}$ \cite{2008EPJST.163..173M} \\
$\alpha$						& 1.84 & $(5.66 \pm 2.67) \times 10^{-6}$ \cite{2008EPJST.163..173M} \\
$\alpha$						& 1.58 & $(-1.5 \pm 2.6 ) \times 10^{-6}$ \cite{2011AA...529A..28A} \\
$\mu$						& Milky Way & $< 28 \times 10^{-9}$ \cite{2011ApJ...738...26L} \\
$\mu$						& 0.685 & $(-3.5 \pm 1.2) \times 10^{-7}$ \cite{2011ApJ...728L..12K} \\
$\mu$						& 0.89 & $< 9 \times 10^{-6}$ \cite{2012PhRvA..85c2505I} \\
$\alpha^2/\mu$					& Milky Way & $< 3.7 \times 10^{-7}$ \cite{2010AA...516A.113L} \\
$\alpha^2/\mu$					& 5.2 & $< 2 \times 10^{-5}$ \cite{2012AA...540L...9L} \\
$\alpha^2/\mu$					& 2.79 & $(6.9 \pm 3.7) \times 10^{-6}$ \cite{2012ApJ...753..102W} \\
$g_p\alpha^2/\mu$				& 1.17 -- 1.6 &$ (-0.1 \pm 1.3) \times 10^{-6}$ \cite{2012MNRAS.425..556R} \\
$g_p\alpha^2/\mu$				& 1.46 & $(+6.8 \pm 1 (\mathrm{stat}) \pm 6.7$ (max systematic)) $\times 10^{-6}$ \cite{2010ApJ...712L.148K}\\
$g_p\left[\mu\alpha^2\right]^{1.85}$ & 0.247 & $(-1.18 \pm 0.46) \times 10^{-5}$ \cite{2010ApJ...716L..23K} \\
$g_p\left[\mu\alpha^2\right]^{1.57}$ & 0.765 & $(-5.2 \pm 4.3) \times 10^{-6}$ \cite{2012ApJ...746L..16K} \\

\hline
\end{tabular}
\label{Table:Constraints} 
\end{table*}

Astrophysical methods used to acquire constraints on these quantities are based on comparing the differences between the observed line centers of absorption or emission lines from astrophysical sources, to their expected values in a laboratory frame on Earth.
A wide range of different atomic and molecular transitions have been used for this purpose (for a review see \cite{2007A&ARv..14..113G}, and for more recent examples from individual systems in the literature see Table \ref{Table:Constraints}) examples of which include comparisons between the inversion line NH$_3$ and rotational lines from CS and H$_2$CO which resulted in the current best constraints on $\Delta\mu/\mu$ of $-3.5 \times 10^{-7}$ from $z\sim0.685$ to the present \cite{2011ApJ...728L..12K}, whilst limits on the evolution of $\alpha$ of $\Delta\alpha/\alpha = (-1.5 \pm 2.6 ) \times 10^{-6}$ have been set using Si~II and Fe~II absorption lines from a $z=1.5$ Quasar (QSO) \cite{2011AA...529A..28A} and of $\Delta\alpha/\alpha = (-1.7 \pm 1.4 ) \times 10^{-6}$ using the combination of HI and OH lines from a $z=0.765$ absorption system \cite{2012ApJ...746L..16K}.  Evidence for a spatial variation in the fine structure constant has also been presented  \cite{2011PhRvL.107s1101W,2012MNRAS.422.3370K}, with the variation fitting a dipole with a significance at the 4.2$\sigma$ level, in the direction right ascension $17.5\pm0.9$ hours, declination $-58\pm9$ degrees.

In general, using multiple transitions from the same species is preferable to contending with different species, as long as the different transitions have differing dependencies on $\alpha$, as in the former case the lines are more likely to arise from the same spatial location, and thus be less affected by random Doppler shifts of their line positions.  Such `Doppler noise' could give rise to a velocity offset anywhere from $\sim$ 10 km s$^{-1}$, comparable to the typical velocity dispersion of the interstellar medium in galaxies (e.g. \cite{1978ARA&A..16..103V}) up to $\sim$ 100 km s$^{-1}$ if the two species are located in different regions of the same galaxy.

In \cite{2008A&A...479..719L} a method was proposed to compare the redshifts of far infrared (FIR) fine-structure lines and low-lying rotational transitions of the CO molecule.  This combination is sensitive to variations in the combined quantity $F=\alpha^2/\mu$. 

The advantage of FIR fine-structure lines is that they are approximately 30 times more sensitive to variations in $\alpha$ than optical and UV transitions.  In particular, the fine structure transition of singly ionized carbon $^2P_{3/2} \to ^2P_{1/2}$ has the combined benefit of providing excellent sensitivity as well as intrinsically being one of the brightest emission lines found in star forming galaxies (e.g. \cite{1985ApJ...291..755C,1991ApJ...373..423S}).  

CO transitions then provide a second independent value for the redshift of the source.  The frequencies of these emission lines are dependant on $\mu$ such that by calculating the quantity $\Delta z = z_{\mathrm{rot}} - z_{\mathrm{fs}}$ leads to the combined quantity $F$:

\begin{equation}
\frac{\Delta z}{1+z_{avg}} = \frac{\Delta F}{F},
\end{equation}
where $z_{\mathrm{avg}}$ is the weighted average of available redshift measurements.

This method has been used for high redshift sources to place constraints on $\Delta F/F$.  For example in a $z=5.2$ system \citet{2012AA...540L...9L} place limits of $\Delta F/F <  2\times 10^{-5}$ using the CO~(J = $7\to6$) rotational transition combined with the [C~I] $^3P_2$ $\to$ $^3P_1$ fine structure transition.

Here we apply this method to the analysis of three systems;  Using the CO~(J = $2\to1$) rotational transition and the [C~II] $^2P_{3/2} \to$ $^2P_{1/2}$ fine structure transition from the $z$=6.4 QSO J1148+5251 and from the QSO BR1202$-$0725 and its companion sub-millimetre galaxy (SMG) using the CO~(J = $5\to4$) rotational transition and the above [C~II] line we derive values of  $\Delta F/F$.

We do this firstly following a traditional approach where we fit a single parameterized model to both emission lines for a given system and compare the line centroids to determine values of $\Delta z$, however due to the very high signal to noise in some of the observations, it becomes increasingly possible, and important, to account for the non-Gaussian features in a line profile.  We therefore take the approach of modelling the line profiles using shapelets, which provide a powerful way of describing any physical line shape by expressing it in terms of a particular orthonormal basis.  We thus take the approach that, whilst the emission lines from a particular system might not be perfectly Gaussian, if the emission is co-located spatially, then they should share the same non-Gaussian traits.  

We also present a second, non-parametric method of analyzing the spectral data;  performing our analysis using the principles of Bayesian inference, we are able to not only quantify objectively the probability that two emission lines from the same source share the same line shape, but by marginalizing over the coefficients for all possible models, we calculate the posterior probability distribution for $\Delta F/F$ independent of any particular choice of model.

In Section \ref{section:Observations} we describe the observations used in this analysis.  In Section \ref{section:Shapelets} we describe the basic mathematical framework behind shapelets, their application to our data, and the marginalization process.  We present our results from all three systems in Section \ref{section:results}, and our discussion of these results, along with our conclusions in Sections \ref{section:Discussion} and \ref{section:Conclusions} respectively.

\section{Observations}
\label{section:Observations}

The observations of the [C~II] line for J1148+5251 were made with the Plateau de Bure Interferometer (PdB) in 2011\cite{2012MNRAS.425L..66M} with a total on source integration time of 17.5 hours.  We use channels smoothed to a width of 72 MHz with a noise level per channel of 0.79 mJy beam$^{-1}$.  An aperture with diameter 4 arc seconds was used to extract the spectra, resulting in a noise level of 1.58 mJy per channel and a total signal to noise ratio for the line of $\sim$ 15.
The CO~(J = $2\to1$) line was observed with the Jansky Very Large Array (JVLA) in 2011.  Channel width was set to 8 MHz, and the noise per channel of 0.036 mJy beam$^{-1}$ resulted in a total signal to noise ratio for the line of $\sim$ 9.  Both emission lines were imaged in a heliocentric velocity frame, however in order to present all the observations in this work in a single velocity frame we convert to LSR velocities by subtracting 10.27 km s$^{-1}$.

For a detailed description of the [C~II] observations made of the two BR1202$-$0725 sources see \cite{2012ApJ...752L..30W}.  To summarise; observations of the two lines were made with the Atacama Large Millimeter Array (ALMA) during 2012.  The observations lasted 25 minutes and produced signal to noise ratios of $\sim$ 80 for both lines, an order of magnitude better than any previous sub-mm line or continuum observation of the system.  The data is binned in channels $\sim$ 16 MHz wide for the QSO, and 32 MHz wide for the SMG, with noise levels per channel of 0.65 and 0.8 mJy beam$^{-1}$ per channel respectively.

The CO~(J = $5\to4$) observations were made with the PdB in February 2007, for a detailed description refer to \cite{2012arXiv1207.6722S}. The channel width in both cases is $\sim$ 30 MHz, with noise per channel of 0.52 mJy, resulting in signal to noise ratios of $\sim$ 14 for both the QSO and SMG.

The PdB observations were imaged in the LSR velocity frame, while the ALMA [C~II] observations were made in a topocentric frame of reference, we therefore add 29 km s$^{-1}$ to the [C~II] observations in order to account for the difference between the two.

\section{Shapelets}
\label{section:Shapelets}

\begin{figure}
\includegraphics[width=80mm]{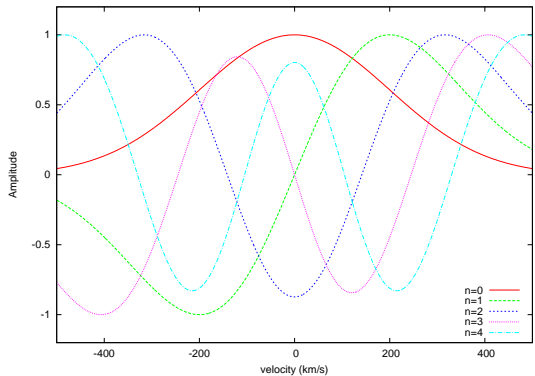} 
\caption{The first 5 1-dimensional shapelet basis functions $B_n$.  The scaling factor is set to 200 km s$^{-1}$, with amplitudes normalised such that each component has a maximum at $1$, or minimum at $-1$.  Note the maxima/minima move further from the centre as $n$ increases.}
\label{figure:shapebasis}
\end{figure}

A thorough description of the Shapelet formalism can be found in \cite{2003MNRAS.338...35R}, with astronomical uses being described in e.g, \cite{2004AJ....127..625K,2005AJ....129.1287K,2003MNRAS.338...48R}.  Here we give only an outline to aid later discussion.

Shapelets are described by a set of dimensionless basis functions, which in one dimension can be written:

\begin{equation}
\phi_n(x) \equiv \left[2^n\sqrt{\pi}n!\;\right]^{-1/2} H_n(x)\;\mathrm{e}^{-x^2/2}
\end{equation}
where $n$ is a non negative integer, and $H_n$ is the Hermite polynomial of order $n$.

These are then modified by a scale factor $\beta$ in order to construct the dimensional basis functions:

\begin{equation}
B_n(x;\beta) \equiv \beta^{-1/2} \phi_n(\beta^{-1}x),
\end{equation}
the first 5 basis functions are shown in Fig. \ref{figure:shapebasis}.   These basis functions are orthonormal in that we can write

\begin{equation}
\int_{-\infty}^\infty \; \mathrm{d}x \; B_n(x;\beta)B_m(x;\beta) = \delta_{mn},
\end{equation}
where $\delta_{mn}$ is the Kronecker delta so that we can represent a function $f(x)$ as the sum:

\begin{equation}
f(x) = \sum_{n\mathrm{=0}}^{\infty} a_nB_n(x;\beta),
\end{equation}
where $a_n$ are shapelet coefficients.  In practice, when dealing with discretely sampled data, this feature of orthonormality depends greatly on the positions of the sample points, the physical extent of the sample space with respect to the scaling factor $\beta$ and the maximum number of basis vectors $n_{\mathrm{max}}$ used for reconstruction.  For a discussion of this see \cite{2004MNRAS.354..199B}.  Due to the nature of our fitting process we can calculate the orthonormality of the set of basis vectors used for every model.  We then define the quantity $O_{\mathrm{max}}$ which describes the maximum deviation away from orthonormality for our basis vectors, i.e:

\begin{equation}
O_{\mathrm{max}} = \mathrm{Max}\left[ \left|\;1 - \sum_{i=1}^{N_{\mathrm{d}}} a_i^2 \;\right| , \left|\;0 -  \sum_{i=1}^{N_{\mathrm{d}}} \sum_{j=1,j\ne i}^{N_{\mathrm{d}}} a_ib_j\;\right| \right] \;\;\;\; a,b \in\{n\}
\end{equation}
and disregard any points where $O_{\mathrm{max}} > O_{\mathrm{thresh}}$, where $O_{\mathrm{thresh}}$ is a threshold value for acceptance.  We find that for values between 1 and 10$\%$ neither the individual best fits nor the marginalised result vary by more than $1\%$.  Below 1$\%$ a large fraction of all solutions are rejected, and above 10--20$\%$ the amount of power in non orthogonal components can result in erroneous best fits.  For the following work we therefore set $O_{\mathrm{thresh}}$=2$\%$.

The total flux $F_{\mathrm{tot}}$ is given by

\begin{equation}
F_{\mathrm{tot}} =\;\int_{-\infty}^{\infty} \;\mathrm{d}x \; f(x) \; = \; \sum_n^{n\mathrm{=even}} \; \big[2^{1-n}\pi^{\frac{1}{2}}\beta\big]^{\frac{1}{2}} {n \choose n/2}^{\frac{1}{2}},
\end{equation}
and the r.m.s deviation of the function $f(x)$, $\sigma_{\mathrm{m}}$ is given by

\begin{eqnarray}
\sigma^2_{\mathrm{m}} &=& \;\int_{-\infty}^{\infty} \;\mathrm{d}x \; x^2f(x) \nonumber \\
                                       &=& \pi^{\frac{1}{4}}\beta^{\frac{5}{2}} F^{-1}\sum_n^{n\mathrm{=even}} \; 2^{\frac{1}{2}(1-n)}a_n\;(1+2n)\;{n \choose n/2}^{\frac{1}{2}}.
\end{eqnarray}

We use this definition of the rms radius to define the full width at half maximum (FWHM) of a particular line shape as 2.3548$\sigma_{\mathrm{m}}$.

\subsection{Model Selection}
\label{section:modelselection}

By fitting a model to an observed spectrum using a set of $n$ shapelet coefficients, we would like to be able to make model dependent inferences about the parameters of that spectrum, a task for which Bayesian Inference is ideal.  Using the MULTINEST algorithm \cite{2009MNRAS.398.1601F}, we are able to analyse efficiently spectral data containing individual, or multiple emission lines, where the Bayesian evidence returned by the algorithm can be used to objectively perform model selection between both different numbers of coefficients, and their values.

We have the following scenarios:
\begin{enumerate}
\item
We have a two emission lines and fit two independent models using the parameters $\nu_i$, $\beta_i$, $C_i$ and a set of $n_i$ shapelet coefficients.  Here $\nu$ is the central frequency, $\beta$ is the scaling parameter, $C$ is a continuum component, $n$ represents the set of shapelet coefficients used in the model and the subscript $i=1,2$ refers to the particular emission line. 
\item
We have two emission lines and fit a joint model using the parameters $\nu_1$, $\nu_2$, $\beta$, $C$, $\alpha$ and a set of $n$ shapelet coefficients.  Here $\nu_1$ and $\nu_2$ are the central frequencies of the two emission emission lines which are allowed to vary independently, whilst $\beta$, the scaling parameter, $C$, a continuum component, and $n$, the set of shapelet coefficients used in the model are the same for both emission lines.  Finally $\alpha$ is an amplitude scaling parameter representing the ratio of the two line peaks.
\end{enumerate}
By comparing the Bayesian evidence for these two scenarios we are therefore able to objectively determine whether the two line shapes are consistent with one another within the limits provided by the data.
In both these cases we also determine values of the redshift associated with the central frequencies of the emission lines, and in those cases where we fit the joint model we calculate $\Delta z = z_{\mathrm{CO}} - z_{\mathrm{CII}}$,  and $\Delta F/F$ for the model, and the errors associated with these quantities.  

The only constraint that we place on the line shapes produced by the shapelet coefficients is that they must be positive for all values of velocity.  We therefore reject all points that have any negative amplitudes by subtracting a large value from their log likelihood when a negative amplitude is present. 

\subsection{Marginalizing over all Possible Line Shapes}

Whilst the advantages of a Bayesian approach are clear, in that we can robustly determine both the maximum number of model components supported by the data, and the coefficients that return the greatest Bayesian evidence, we are also able to marginalise over all the line shapes described by our shapelet coefficients in order to calculate the probability distribution of $\Delta z$ alone.

The probability of a particular value of $\Delta z$ resulting from model parameters $\theta$ is given by:

\begin{equation}
P(\Delta z, \theta | D) \propto \exp\big[-0.5(D - M)^TN^{-1}(D - M)\big],
\end{equation}
where $D$ is our data vector, $M$ is the vector containing the model described by the parameters $\theta$ and $N$ is the covariance matrix, which in all the following work we take to be diagonal, with elements $N_{ii} = \sigma_i^2$ where $\sigma_i$ is the noise in channel $i$.

The probability of a particular value of $\Delta z$ is therefore given by:

\begin{equation}
P(\Delta z) = \int \; P(\theta | D) \mathrm\;{d}\theta,
\end{equation}
where we have integrated over all model parameters $\theta$.  In this way we can account for the uncertainty in choosing a particular model, and may also include systems where the line shape is not well described by a gaussian, as is the case for the SMG in the BR1202 system described in Section \ref{section:SMG}.
 
\section{Results}
\label{section:results}

\subsection{J1148+5251}

\subsubsection{Line Profiles}

\begin{figure*}
\begin{minipage}{168mm}
\begin{center}$
\begin{array}{cc}
\includegraphics[width=80mm]{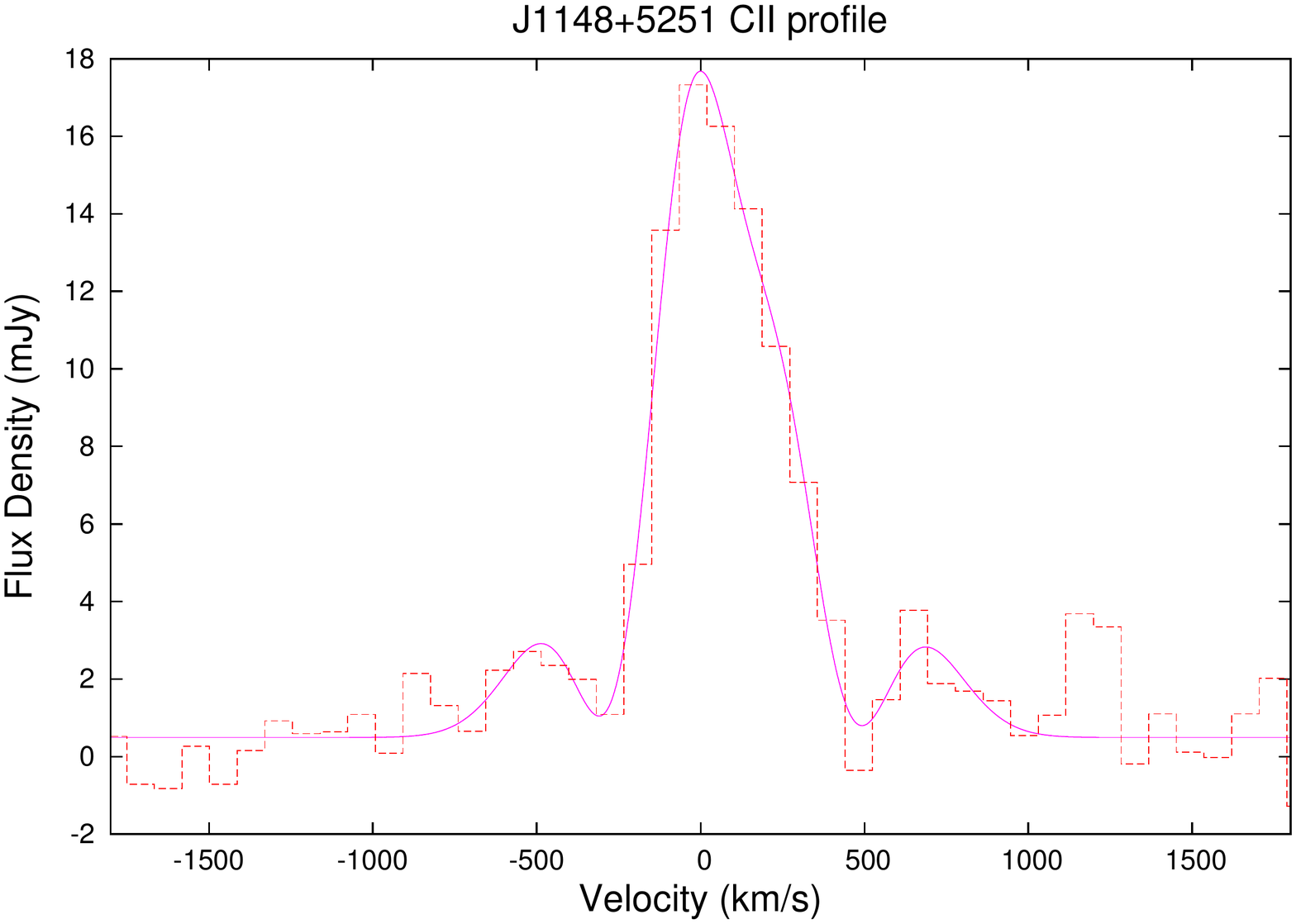} &
\includegraphics[width=80mm]{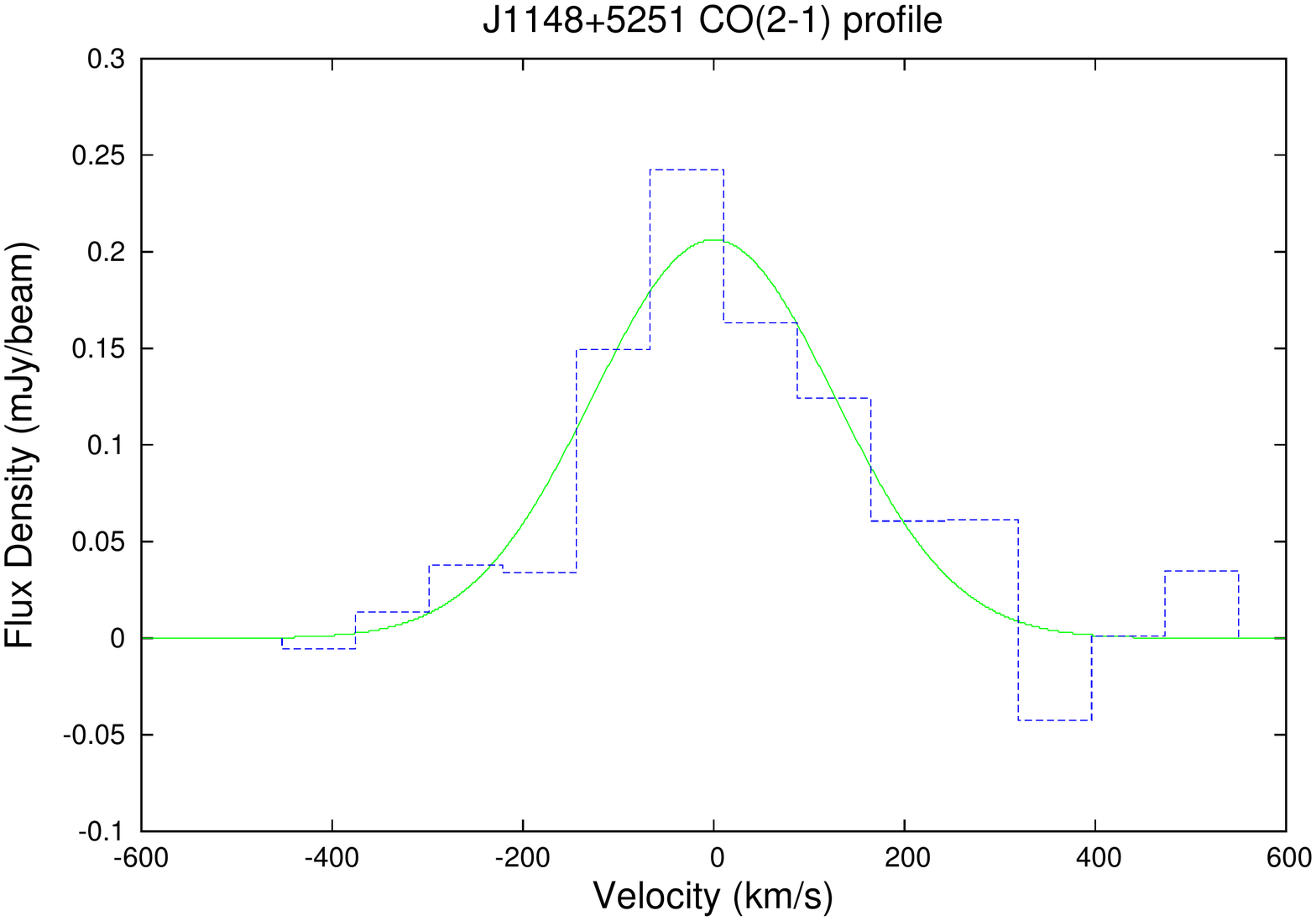}
\end{array}$
\end{center}
\caption{(Left) [C~II] spectrum (red) with the best fit model (magenta) overlaid and (Right) CO~(J=$2\to1$) spectrum (blue) with the best fit model (green) overlaid for J1148+5251.  In both cases zero velocity is chosen to be at the peak of the respective model and not at a specific redshift value.}
\label{figure:1148profiles}
\end{minipage}
\end{figure*}

The [C~II] and CO~(J = $2\to1$) spectra for J1148+5251 are shown in Fig. \ref{figure:1148profiles}.  The evidence for the [C~II] line supports a fit with $n_{\mathrm{max}}$ = 7 indicating large deviations from the $n=0$ term (i.e. a single Gaussian).  This can be seen in the spectrum as a set of extended wings originating from strong quasar outflows \cite{2012MNRAS.425L..66M}.  The central Gaussian component has a FWHM of $306 \pm 20$ km s$^{-1}$, with an integrated flux of 7.5 Jy km s$^{-1}$.  This represents 72$\%$ of the total flux in the emission line, with 2.9 Jy km s$^{-1}$ located in the non Gaussian components.  The total FWHM of the line model is $395 \pm 30$ km s$^{-1}$.  We can compare this line width to the result of fitting only a single Gaussian model, which results in a FWHM of 400 $\pm$  20 km s$^{-1}$. 
In contrast the CO~(J = $2\to1$) line is adequately described by a single component model, with a FWHM of $297 \pm 49$ km s$^{-1}$.  This is therefore consistent within 2$\sigma$ errors with the CII line.

\subsubsection{Performing a joint fit}

\begin{figure*}
\begin{minipage}{168mm}
\begin{center}$
\begin{array}{cc}
\includegraphics[width=80mm]{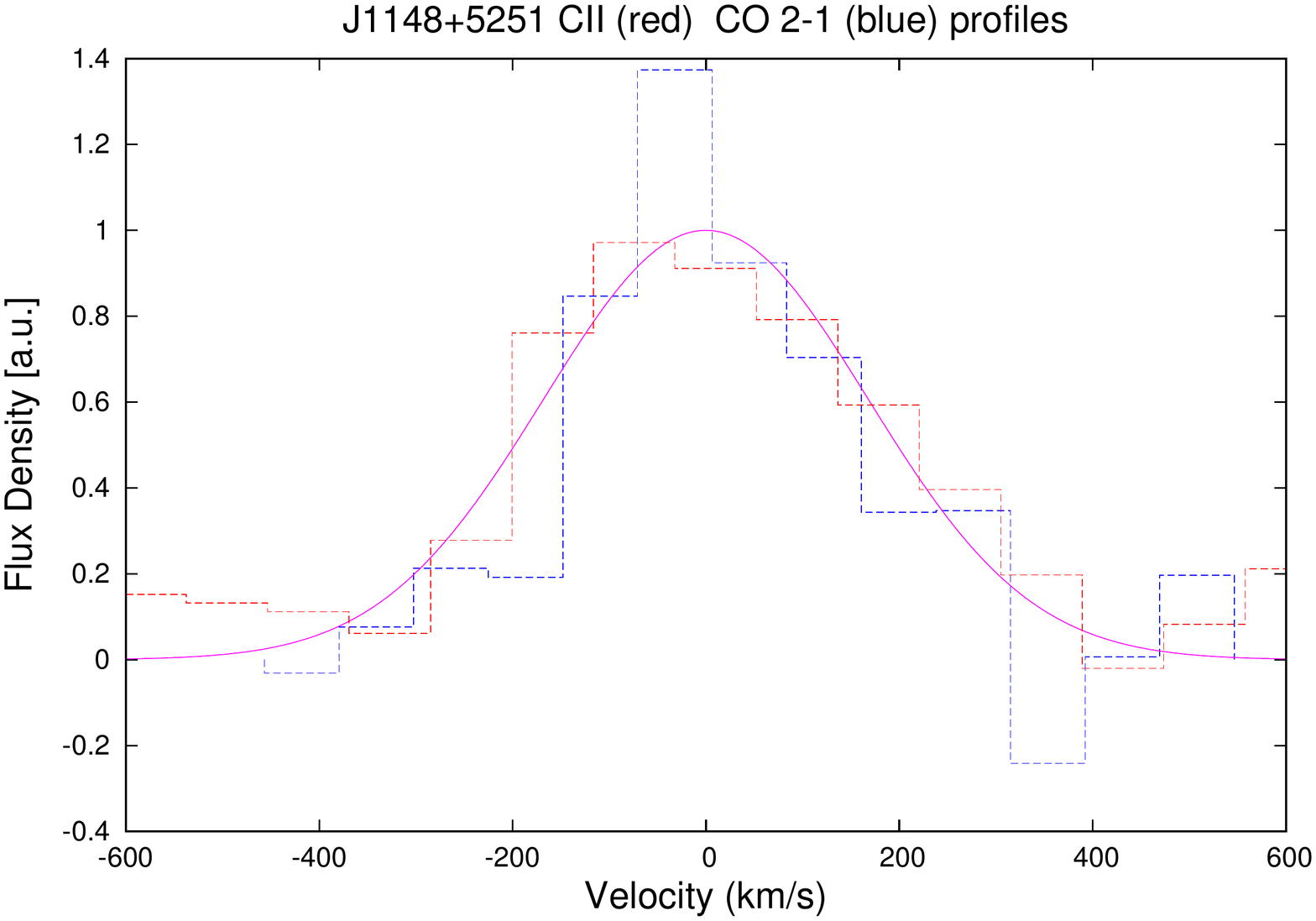} &
\includegraphics[width=80mm]{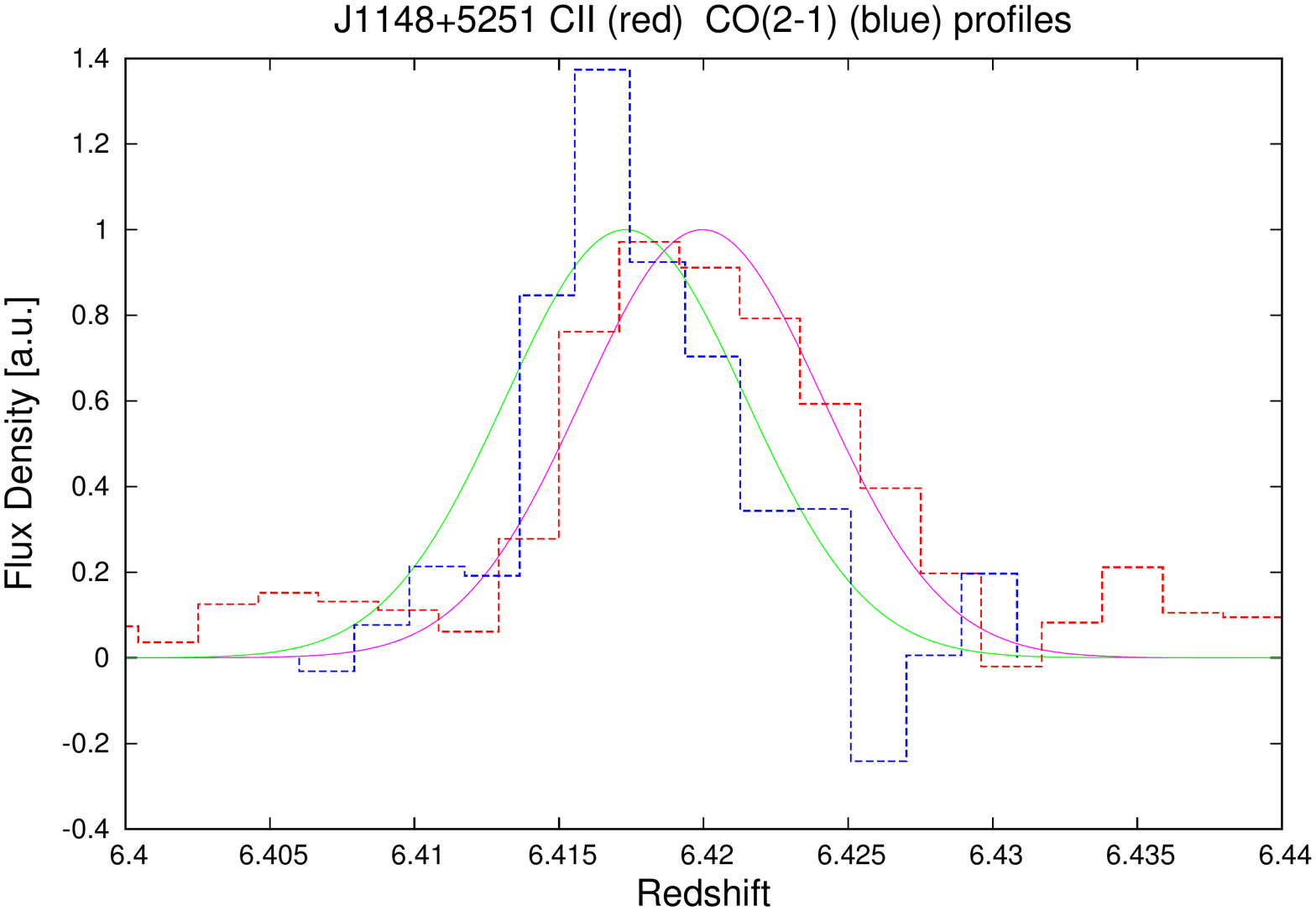}
\end{array}$
\end{center}
\caption{J1148+5251: (Left) The best fit joint fit profile (magenta) overlaid onto both the [CII] (red dashed) and CO~(J= $2\to1$) (blue dashed) emission lines.  Amplitudes have been normalised such that the model has a peak at 1.0, whilst the CO~(J = $2\to1$) and [C~II] emission lines have peaks normalised such that their height relative to the model is as in the best fit.  Both emission lines are shown with zero velocity corresponding to their model redshift.  Right: [C~II] data (red dashed line) and CO~(J = $2\to1$) data (blue dashed line) on a redshift axis with amplitudes normalised as in the left panel.  Redshifts have been calculated using rest frame emission frequencies of 1900.539 GHz and 230.538 GHz for the [C~II] and CO lines respectively.  The solid magenta and green lines are the joint fit model line profiles at their best fit redshift values for the [C~II] and CO~(J = $2\to1$) lines respectively. 
 }
\label{figure:1148jointprofiles}
\end{minipage}
\end{figure*}

\begin{figure*}
\begin{minipage}{168mm}
\begin{center}$
\begin{array}{cc}
\includegraphics[width=80mm]{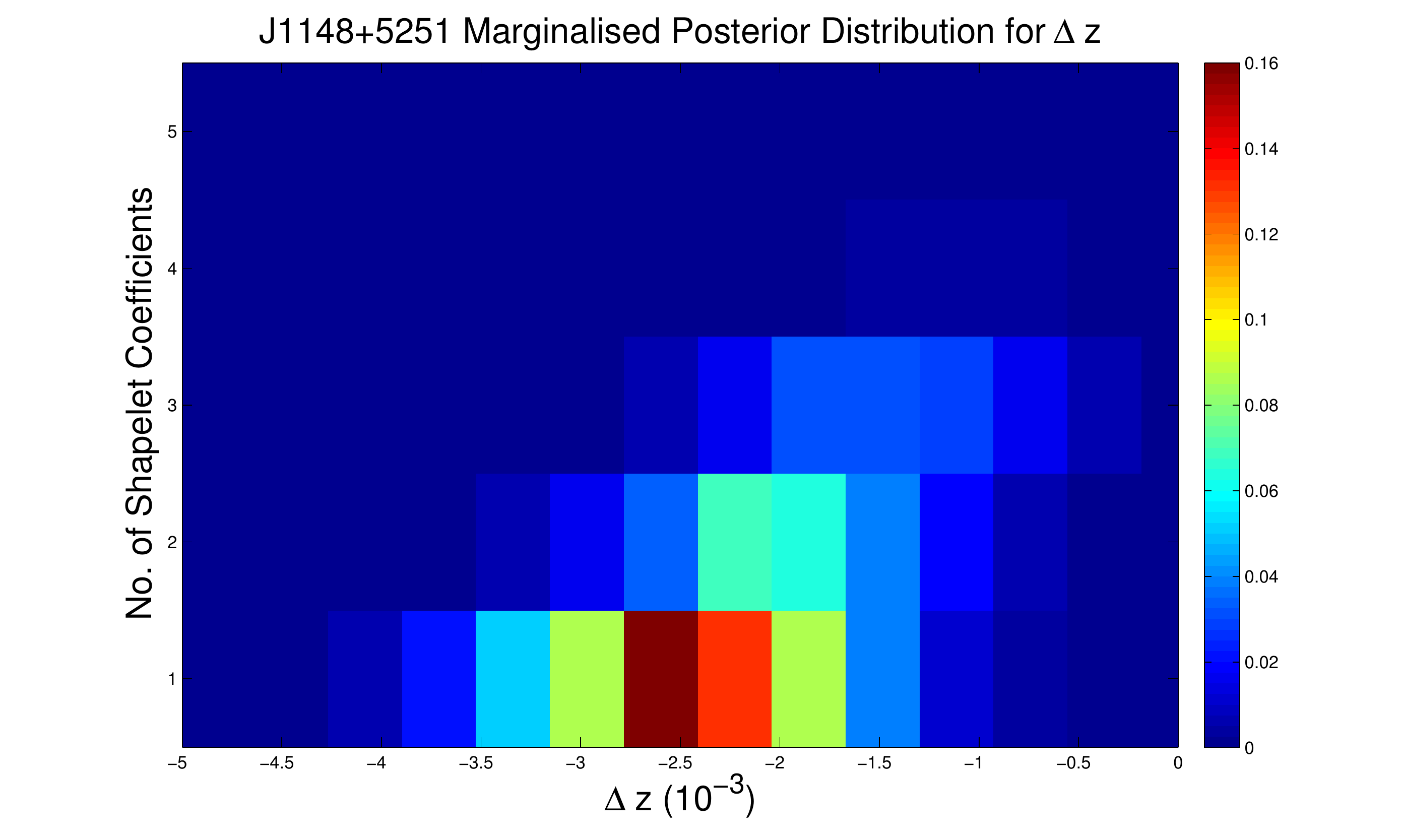} &
\includegraphics[width=80mm]{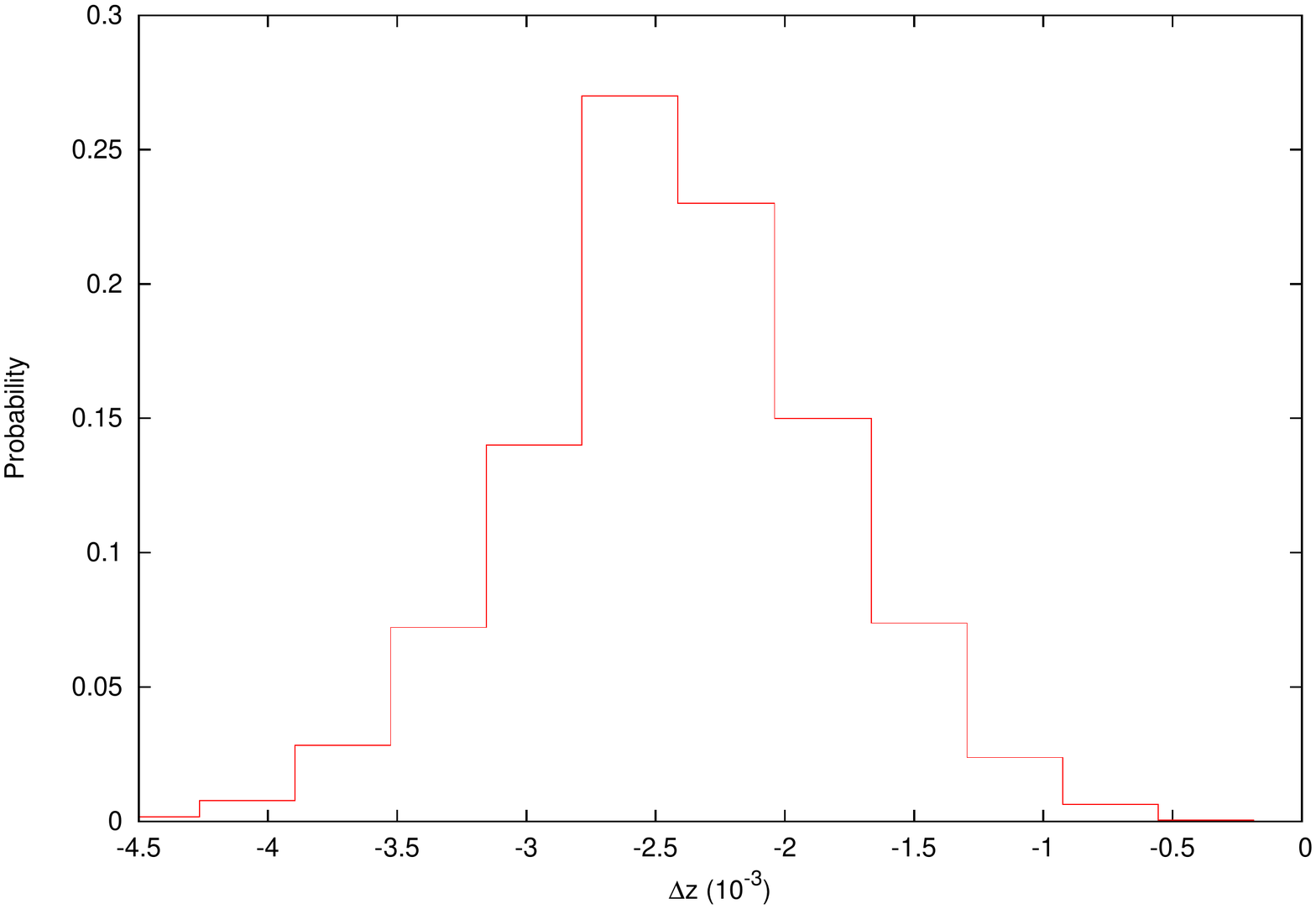}
\end{array}$
\end{center}
\caption{Left: Normalised 2D probability distribution for $n_{\mathrm{max}}$ and $\Delta$ F/F marginalized over shapelet coefficients.  Right: Normalised 1D probability distribution for $\Delta$ F/F marginalized over all $n_{\mathrm{max}}$ and all shapelet coefficients.}
\label{figure:1148marginprofiles}
\end{minipage}
\end{figure*}

As described in Section \ref{section:modelselection} when performing a joint fit we use only the velocity range that is shared between both data sets and as such we only fit for the joint profile out to edge of the CO data (approximately $\pm$ 500 km s$^{-1}$) which means we exclude the majority of the extended wings seen in the [C~II] profile from the fit. 
Fig. \ref{figure:1148jointprofiles} (left) shows the best fit joint fit profile overlaid onto both the [C~II] and CO~(J = $2\to1$) emission lines. 
The consistency between the line shapes can be described quantitively by comparing the Bayesian evidence for a joint fit where the two lines are described by the same set of coefficients, and the evidence when the two lines are allowed to be described by different parameters.  Here the log evidence supports a joint profile with a difference of 1.15, corresponding to a 76$\%$ probability that the two emission lines are described by the same model within the limits of the data. 

\begin{table}
\centering
\caption{J1148+5251 Emission Line Parameters from Joint Fit} 
\centering 
\begin{tabular}{c c c} 
\hline\hline 
Line & $\nu_{\mathrm{obs}}$ & $z$ \\[0.5ex] 
	&	GHz			& \\
\hline 
[C~II] & 256.137 $\pm$ 0.010 & 6.4200 $\pm$ 0.0003 \\
CO~(J = $2\to1$) & 31.081 $\pm$ 0.002 & 6.4173 $\pm$ 0.0005 \\
\hline
\end{tabular}
\label{Table:1148joint} 
\end{table}

The joint fit is described by $n_{\mathrm{max}}$ = 1 and FWHM of $395 \pm 24$ km s$^{-1}$.  Line centroids and redshifts are given in Table \ref{Table:1148joint}.  We find $\Delta z = (-2.7 \pm 0.6 \times 10^{-3})$, which, if the difference in redshifts were due only to changes in the fundamental constants and not the result of intrinsic differences in the line shape would correspond to $\Delta F/F = ( -3.6 \pm 0.8 ) \times 10^{-4}$, representing a 4.5$\sigma$ deviation from zero.  The separation in redshift is shown in Fig. \ref{figure:1148jointprofiles} (Right).

Fig \ref{figure:1148marginprofiles} shows the results of marginalizing over all model line profiles.  The left panel shows the normalised 2D probability distribution for $n_{\mathrm{max}}$ and $\Delta z$ marginalized over all shapelet coefficients for that value of $n_{\mathrm{max}}$.  The peak is consistent with a value of $\Delta z \sim -2.4\times10^{-3}$ for all $n_{\mathrm{max}}$ with significant probabilities. Beyond this region the probability distribution is zero to machine precision.  The right panel shows this 2D distribution marginalised over all $n_{\mathrm{max}}$.  The final result is consistent with a 99$\%$ confidence interval for the value $\Delta z$ of $(-2.4 \pm 1.7 ) \times 10^{-3}$.  

\subsection{BR1202$-$0725 QSO component}

\subsubsection{Line Profiles}

\begin{figure*}
\begin{minipage}{168mm}
\begin{center}$
\begin{array}{ccc}
\includegraphics[width=60mm]{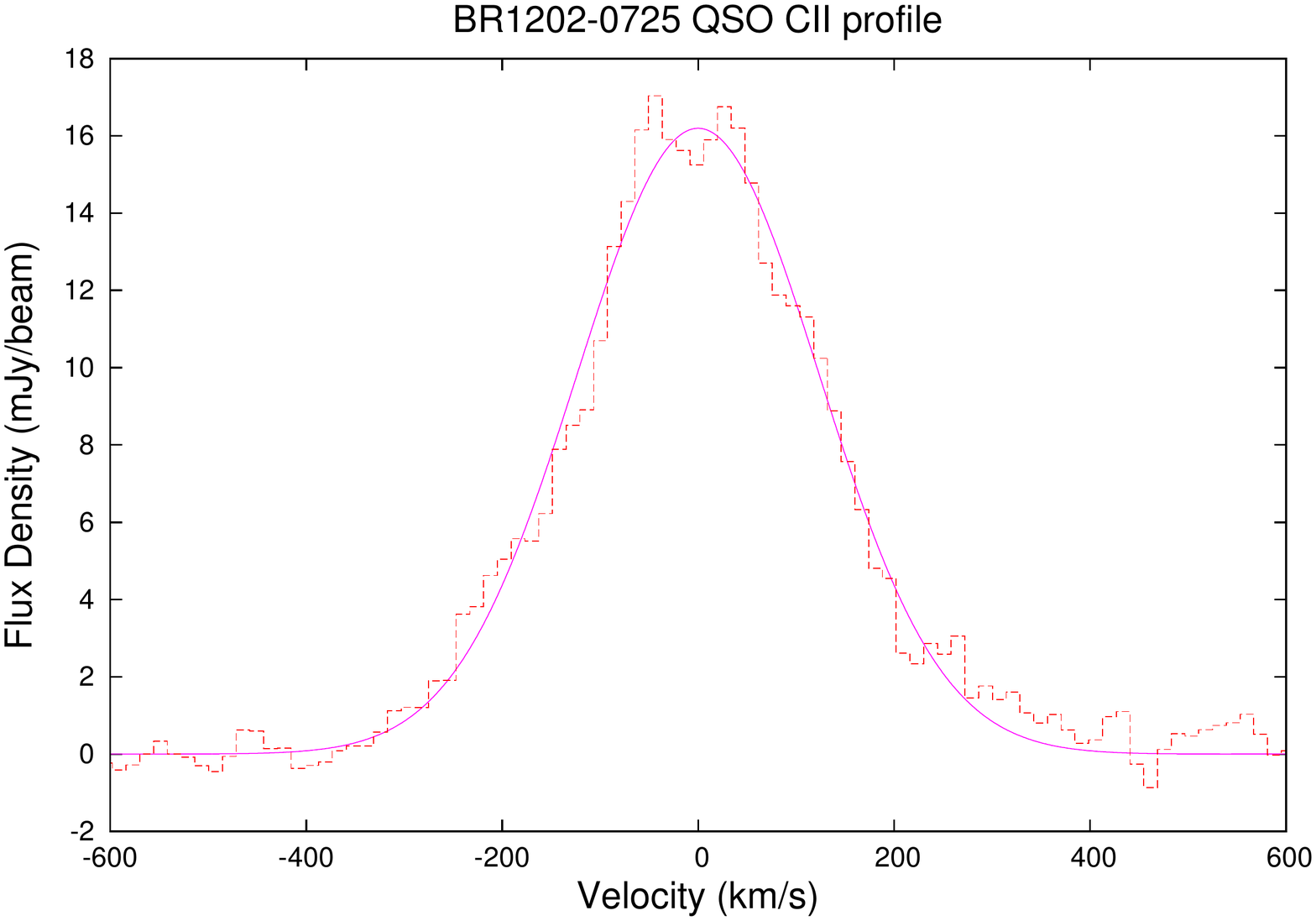} &
\includegraphics[width=60mm]{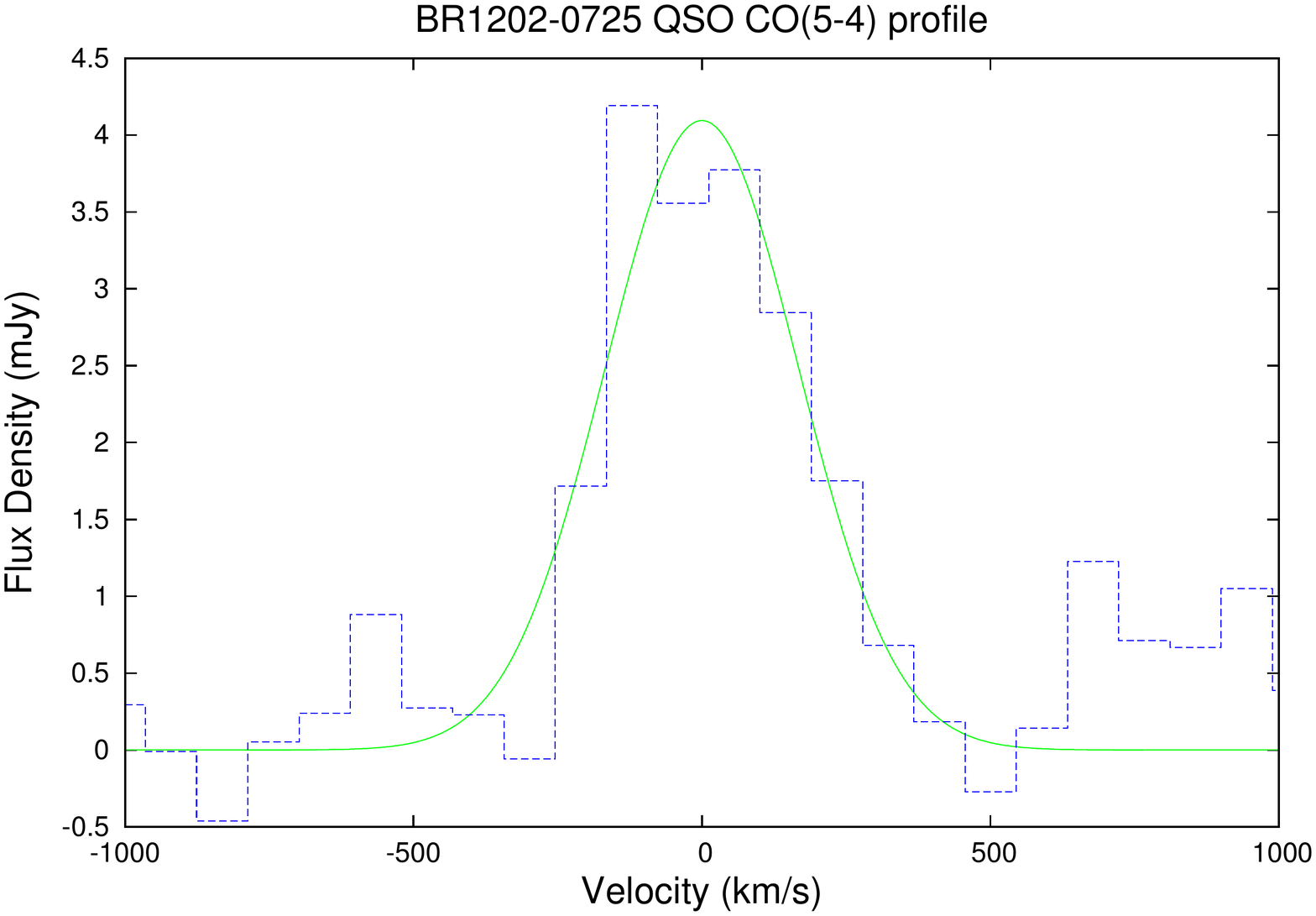}
\end{array}$
\end{center}
\caption{BR1202-0725: (Left) The [C~II] spectrum (red) with the best fit model (magenta) overlaid. (Right) CO~(J = $5\to4$) spectrum (blue) with the best fit model (green) overlaid. In both cases zero velocity is chosen to be at the peak of the respective model and not at a specific redshift value.}
\label{figure:1202QSOprofiles}
\end{minipage}
\end{figure*}

The [C~II] and CO~(J = $5\to4$) spectra for the BR1202 QSO are shown in Fig. \ref{figure:1202QSOprofiles}.  The [C~II] emission line is best modeled by a single Gaussian with a FWHM of 290 $\pm$ 4 km s$^{-1}$, whilst the CO~(J = $5\to4$) line is described by a single gaussian with a FWHM of 390 $\pm$ 30 km s$^{-1}$.  As such the two profiles are not consistent with one another within their errors, with the CO emission line being $\sim$ 30$\%$ broader than the [C~II] line.

\subsubsection{Performing a joint fit}

\begin{figure*}
\begin{minipage}{168mm}
\begin{center}$
\begin{array}{cc}
\includegraphics[width=80mm]{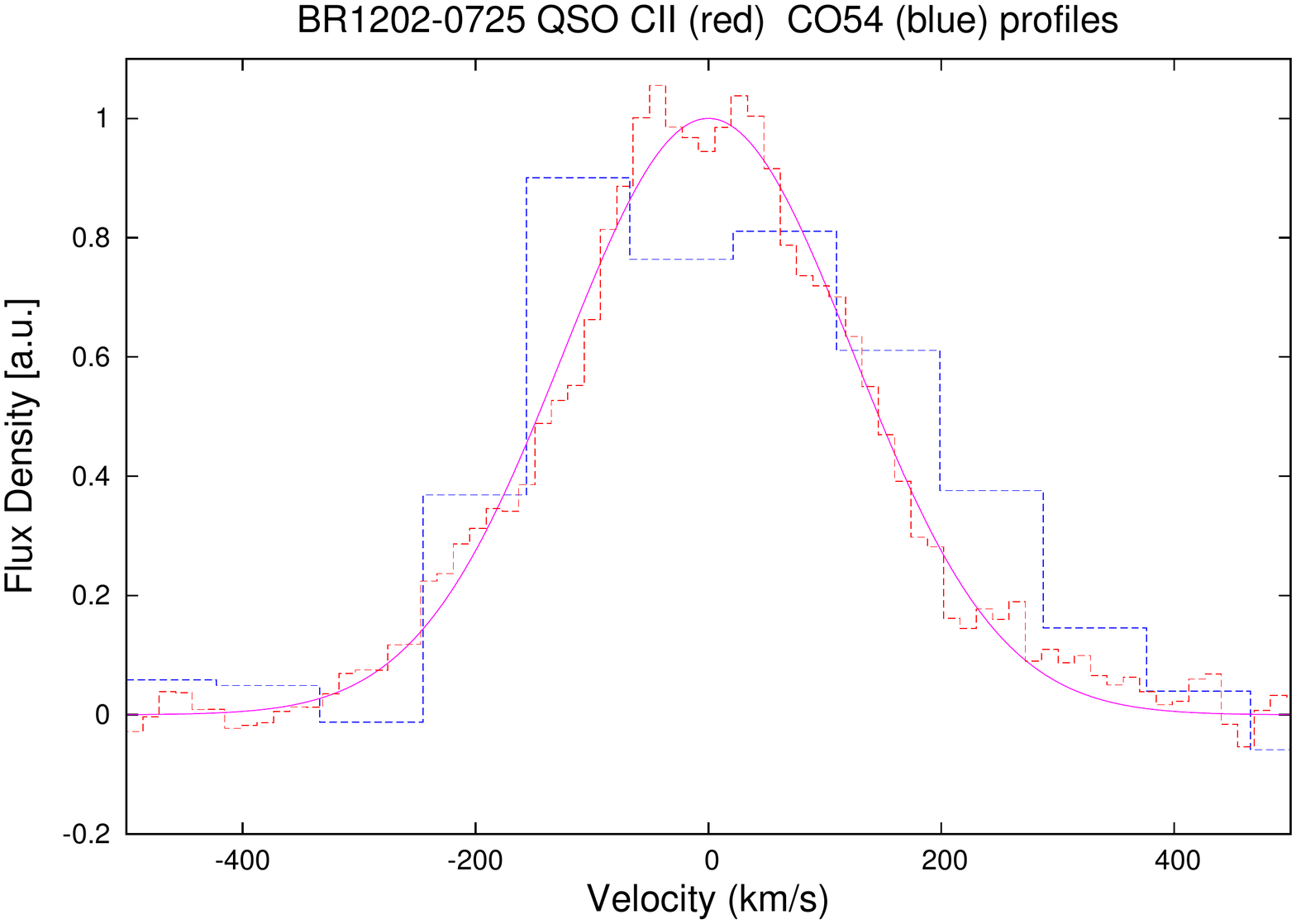} &
\includegraphics[width=80mm]{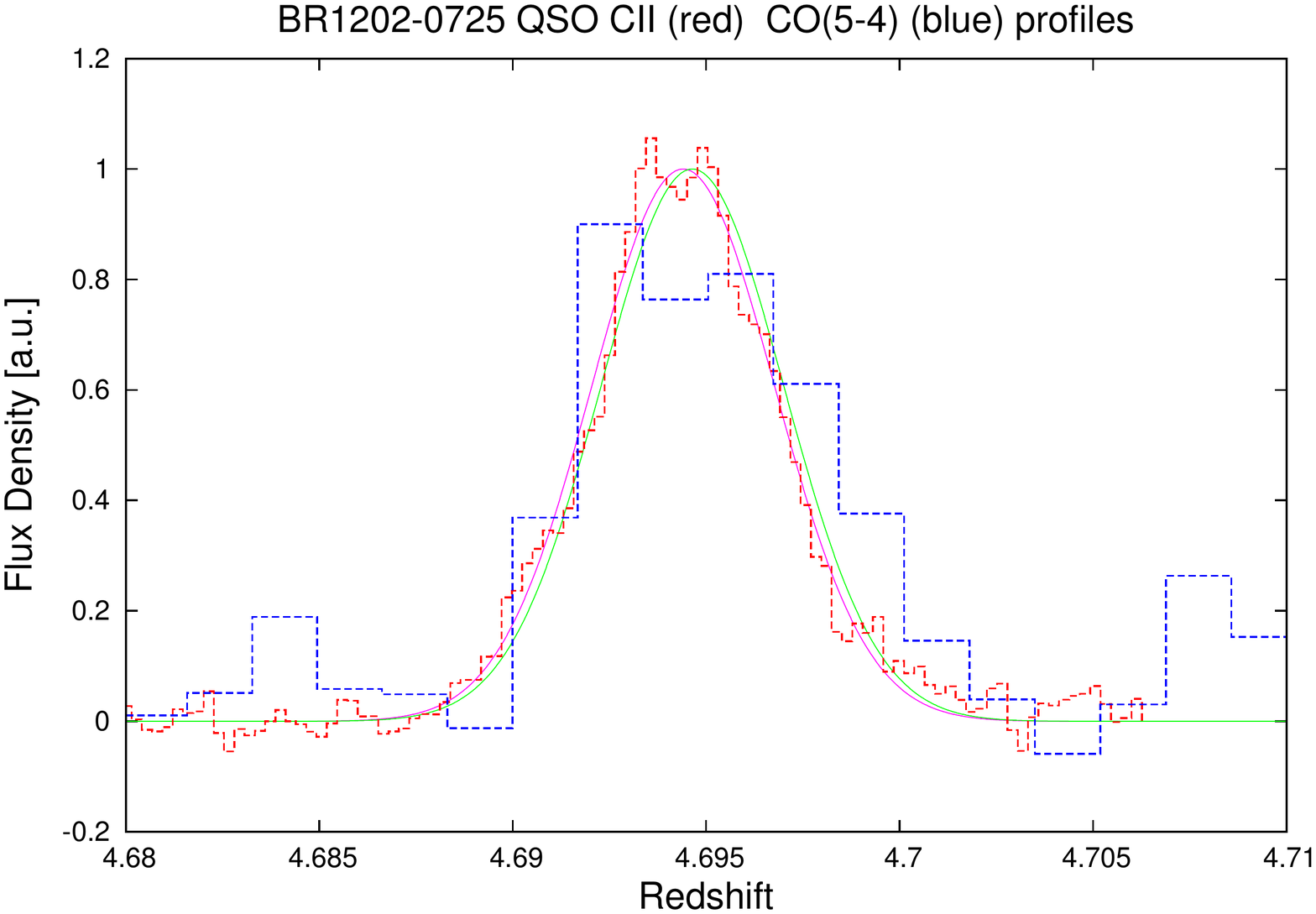}
\end{array}$
\end{center}
\caption{BR1202-0725: (Left) The best fit joint fit profile (magenta line) overlaid onto both the [C~II] (red dotted) and CO~(J = $5\to4$) (blue dotted) emission lines.  Amplitudes have been normalised such that the model has a peak at 1.0, whilst the CO~(J = $5\to4$) and [C~II] emission lines have peaks normalised such that their height relative to the model is as in the best fit.  Both emission lines are shown with zero velocity corresponding to their model redshift. (Right) [C~II] data (red dashed line) and CO~(J = $5\to4$) data (blue dashed line) on a redshift axis with amplitudes normalised as in the left panel.  Redshifts have been calculated using rest frame emission frequencies of 1900.539 GHz and 576.267931 GHz for the [C~II] and CO lines respectively. The solid magenta and green lines are the joint fit model line profile at the best fit redshift values for the [C~II] and CO~(J = $5\to4$) lines respectively. 
 }
\label{figure:1202jointprofiles}
\end{minipage}
\end{figure*}

\begin{figure*}
\begin{minipage}{168mm}
\begin{center}$
\begin{array}{cc}
\includegraphics[width=80mm]{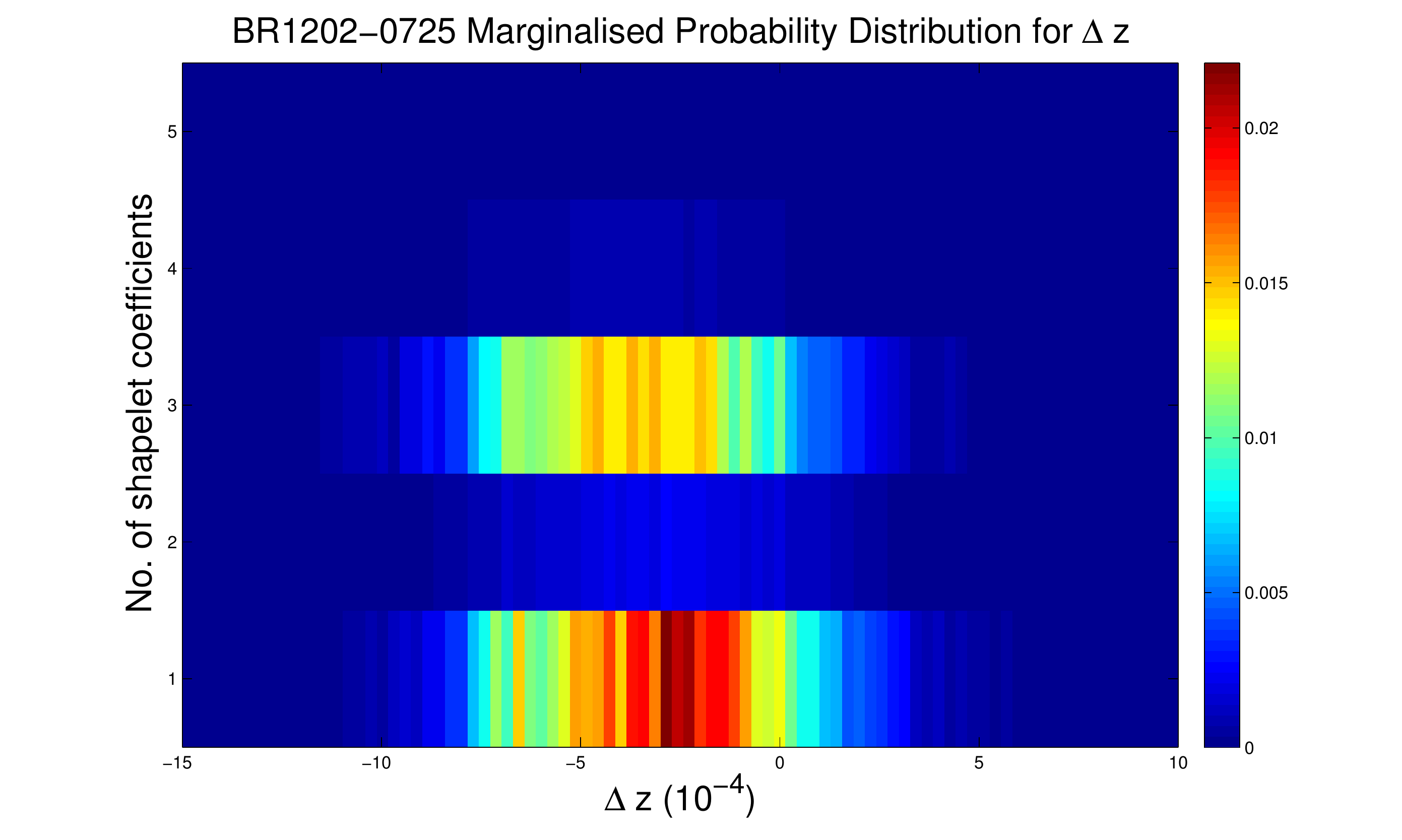} &
\includegraphics[width=80mm]{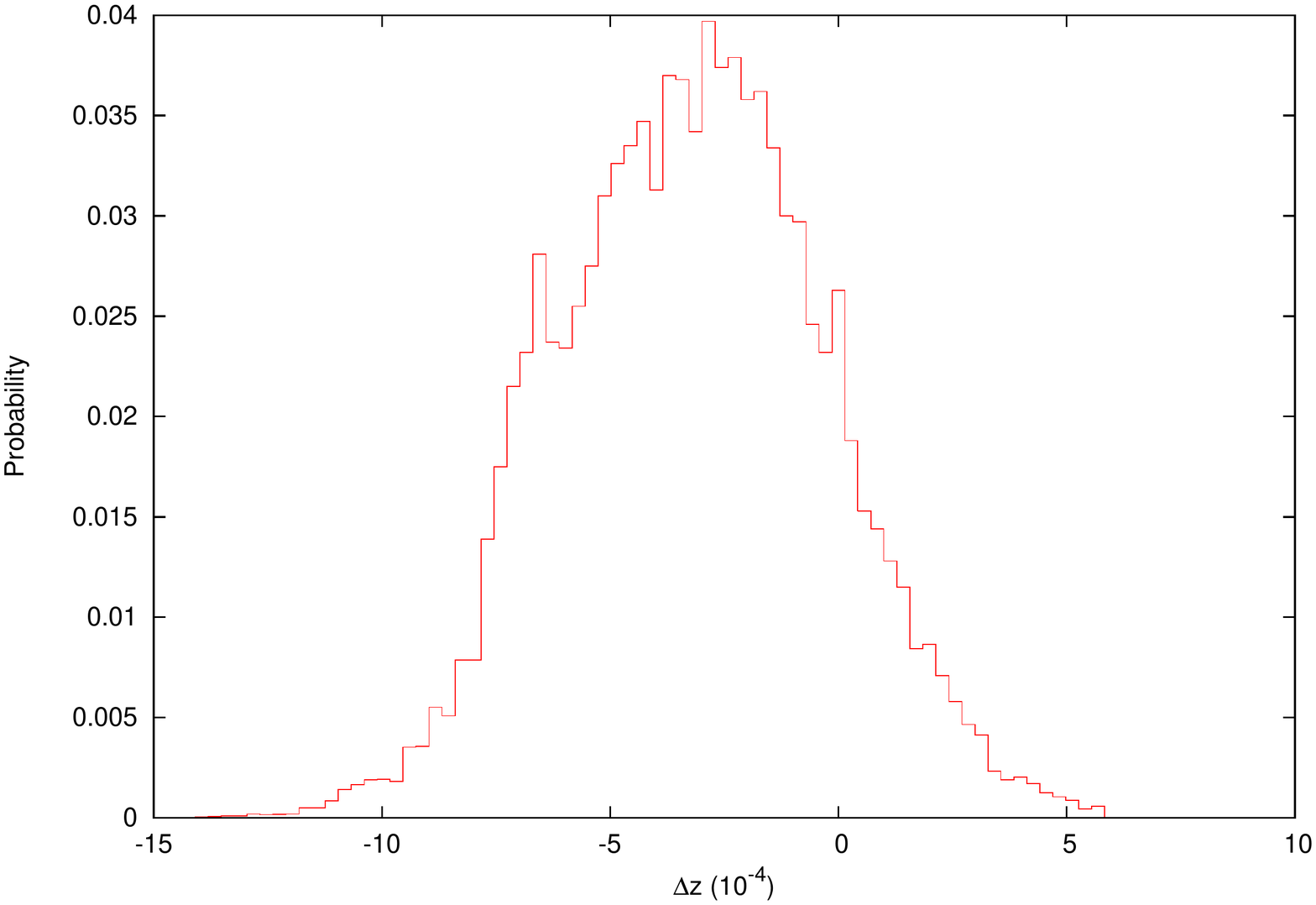}
\end{array}$
\end{center}
\caption{Left: Normalised 2D probability distribution for $n_{\mathrm{max}}$ and $\Delta z$ marginalized over shapelet coefficients.  Right: Normalised 1D probability distribution for $\Delta z$ marginalized over all $n_{\mathrm{max}}$ and all shapelet coefficients.  }
\label{figure:1202marginprofiles}
\end{minipage}
\end{figure*}

The difference in the line widths between the [C~II] and CO~(J = $5\to4$) emission lines is demonstrated in the difference in the evidence values when comparing the joint fit to independent models, with a difference of approximately 0.5, corresponding to a 62$\%$ probability that the two lines do not share the same line profile.
\begin{table}
\centering
\caption{BR1202$-$0725 QSO Emission Line Parameters from Joint Fit} 
\centering 
\begin{tabular}{c c c} 
\hline\hline 
Line & $\nu_{\mathrm{obs}}$ & $z$ \\[0.5ex] 
	&	GHz			& \\
\hline 
[C~II] & 333.723 $\pm$ 0.002 & 4.69496 $\pm$ 0.00003 \\
CO~($5\to4$) & 101.195 $\pm$ 0.005 & 4.6946 $\pm$ 0.0003 \\
\hline
\end{tabular}
\label{Table:1202QSOjoint} 
\end{table}
The joint fit is best described by a single gaussian with FWHM 293 $\pm$ 6 km s$^{-1}$.  Line centroids and redshifts are given in Table \ref{Table:1202QSOjoint}.  From these we find values of $\Delta z = (-3 \pm 3) \times 10^{-4}$, and $\Delta F/F = -5 \pm 5 \times 10^{-5}$.  

Since the evidence has a slight preference for independent models for the two lines, we also calculate values of $\Delta z$ and $\Delta F/F$ for the two independently fitted models in order to see how large an effect under fitting the data with a single model can have on the result.
In this case the redshifts of the two emission lines are given by $z_{\mathrm{CII}}$= 4.69495 $\pm$ 0.00003 and $z_{\mathrm{CO}}$= 4.6948 $\pm$ 0.0003, which gives us $\Delta z = (-1.4 \pm 3) \times 10^{-4}$, and correspondingly $\Delta F/F = (-2 \pm 5) \times 10^{-5}$ if as before we assume this difference in redshift is due only to an evolution in the fundamental constants.  

Finally in Fig. \ref{figure:1202marginprofiles} we show the marginalized posterior distribution $\Delta z$ as for J1148+5251, from which we derive a $99\%$ confidence interval for $\Delta z$ of ( $-3\pm8$) $\times$ 10$^{-4}$.

\subsection{BR1202$-$0725 SMG component}
\label{section:SMG}

\begin{figure*}
\begin{minipage}{168mm}
\begin{center}$
\begin{array}{ccc}
\includegraphics[width=80mm]{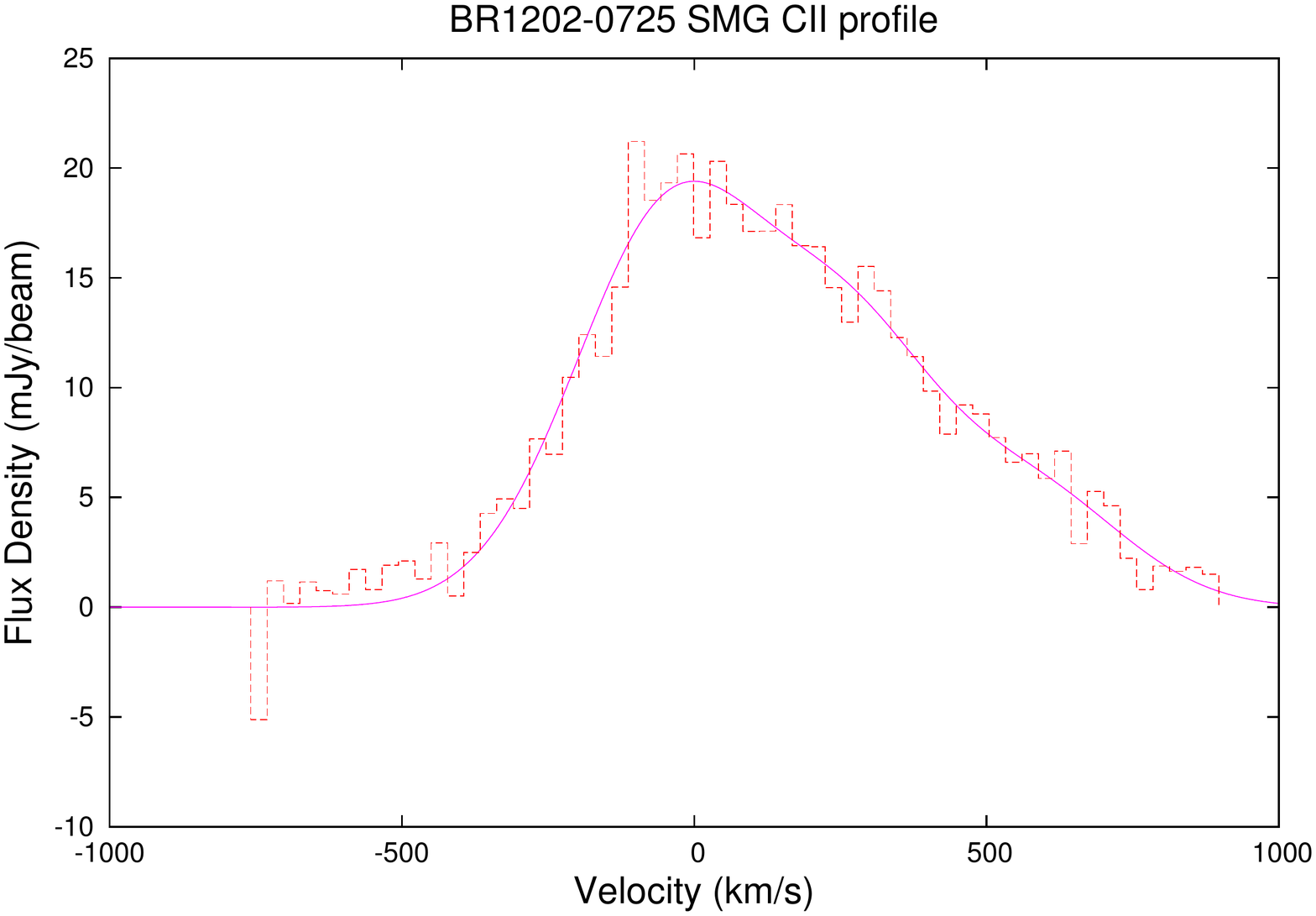} &
\includegraphics[width=80mm]{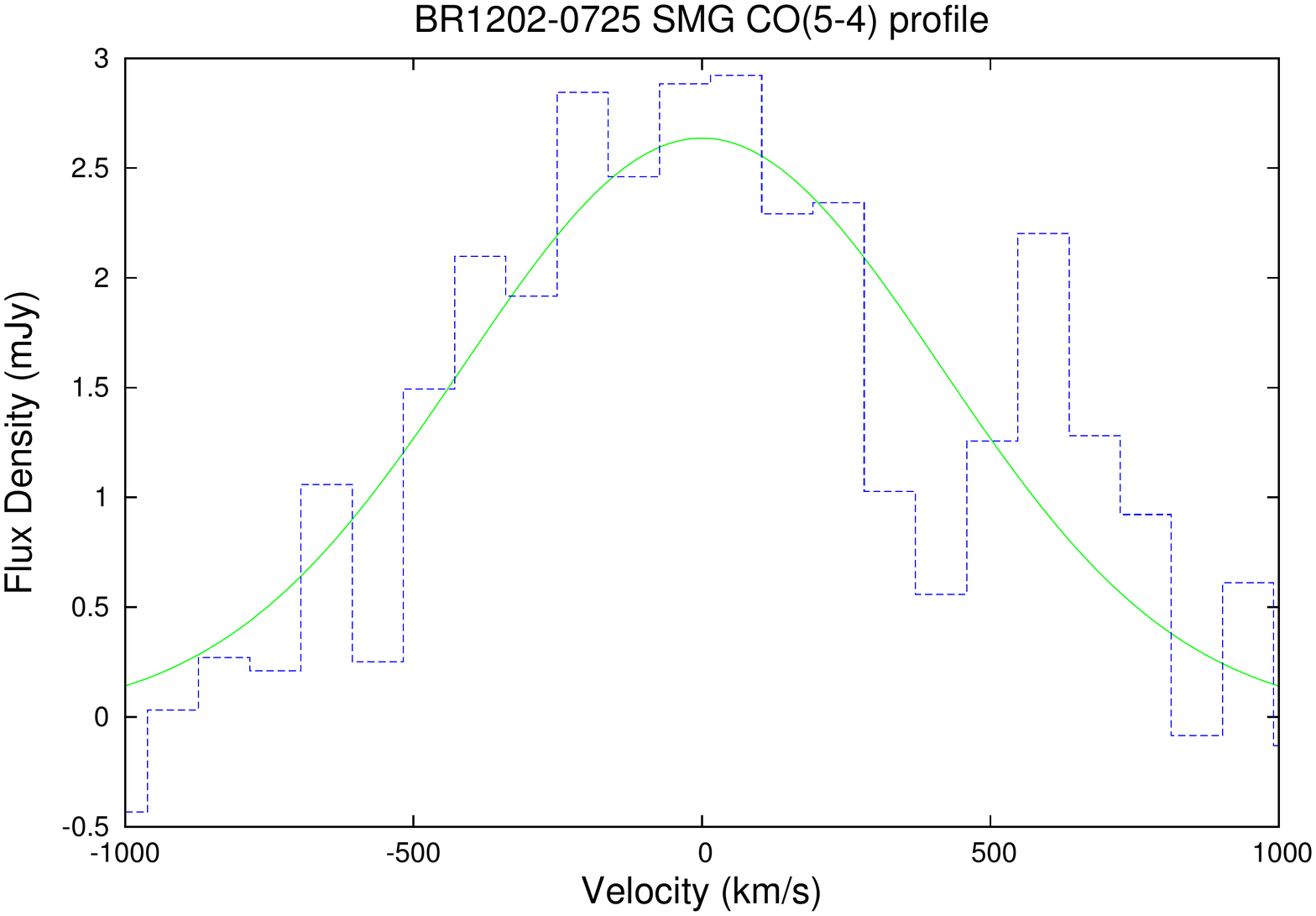}
\end{array}$
\end{center}
\caption{Left: The [C~II] spectrum (red dashed) with the best fit model (solid magenta) overlaid. Right: CO~($5\to4$) spectrum (blue dashed) with the best fit model (solid green) overlaid.  In both cases zero velocity is considered to be at the peak of the model.}
\label{figure:1202SMGprofiles}
\end{minipage}
\end{figure*}

\subsubsection{Line Profiles}

The [C~II] and  CO~(J = $5\to4$) spectra for the BR1202$-$0725 SMG are shown in Fig. \ref{figure:1202SMGprofiles}.  Whilst the previous line profiles have all been roughly Gaussian in nature, the [C~II] emission line has a much more perturbed shape, with $n_{\mathrm{max}}$ of 5 required to adequately describe it with FWHM of $700 \pm 21$ km s$^{-1}$.  The CO line however is simply described by a single Gaussian with FWHM $970 \pm 90$ km s$^{-1}$, such that as with the QSO the  CO~($5\to4$) line is approximately 30$\%$ broader than the [C~II].

\subsubsection{Performing a joint fit}

\begin{figure*}
\begin{minipage}{168mm}
\begin{center}$
\begin{array}{cc}
\includegraphics[width=80mm]{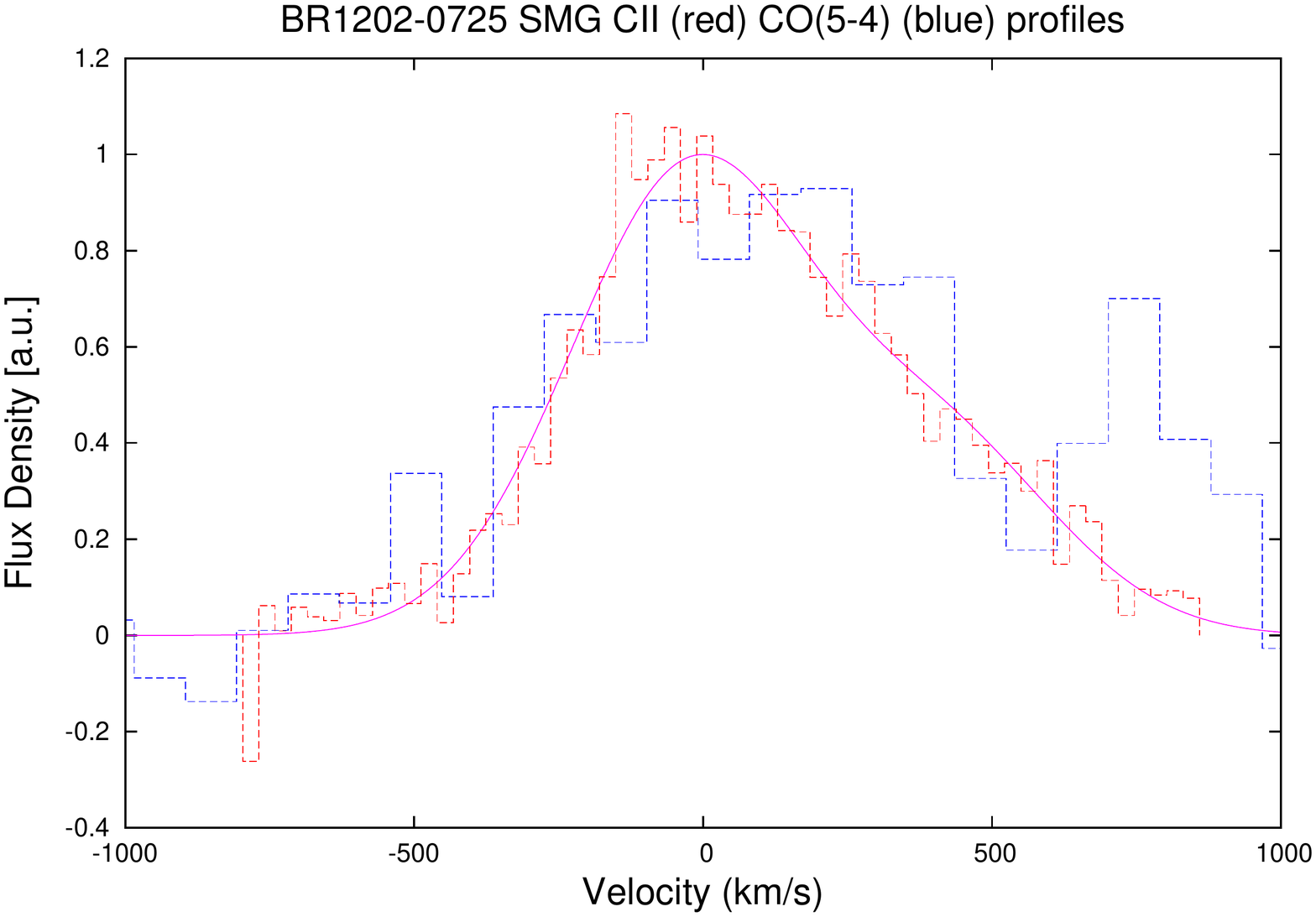} &
\includegraphics[width=80mm]{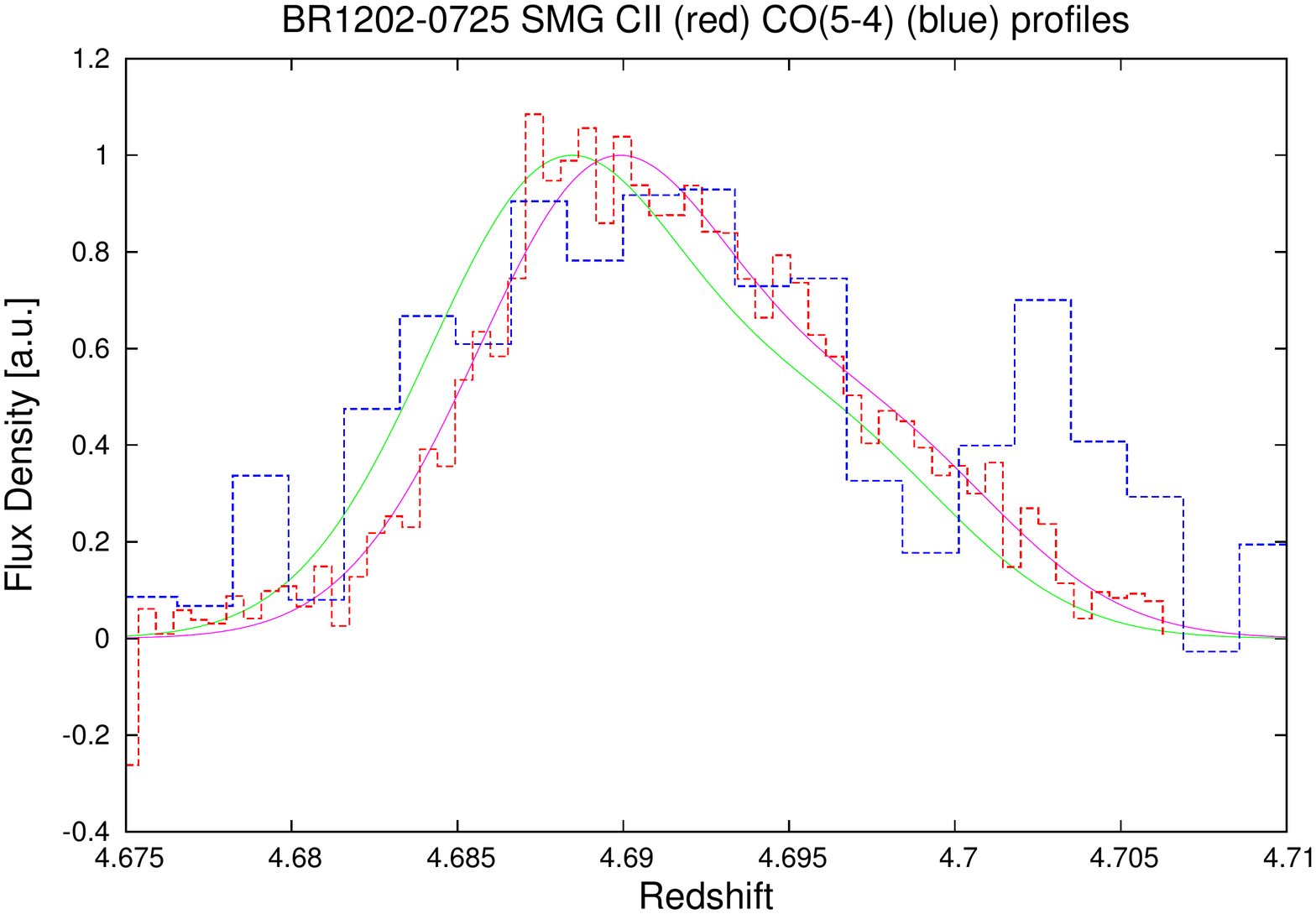}
\end{array}$
\end{center}
\caption{Left: The best fit joint fit profile (solid magenta) overlaid onto both the C~II (red dashed) and  CO~($5\to4$) (blue dashed) emission lines.  Amplitudes have been normalised such that the model has a peak at 1.0, whilst the CO5$\to$4 and CII emission lines have peaks normalised such that their height relative to the model is as in the best fit.  Right: C~II data (red dotted line) and  CO~($5\to4$) data (blue dotted line) on a redshift axis with amplitudes normalised as in the left panel.  The solid magenta and green lines are the joint fit model line profile at the best fit redshift values for the C~II and CO~($5\to4$) lines respectively. 
 }
\label{figure:1202SMGjointprofiles}
\end{minipage}
\end{figure*}

\begin{figure*}
\begin{minipage}{168mm}
\begin{center}$
\begin{array}{cc}
\includegraphics[width=80mm]{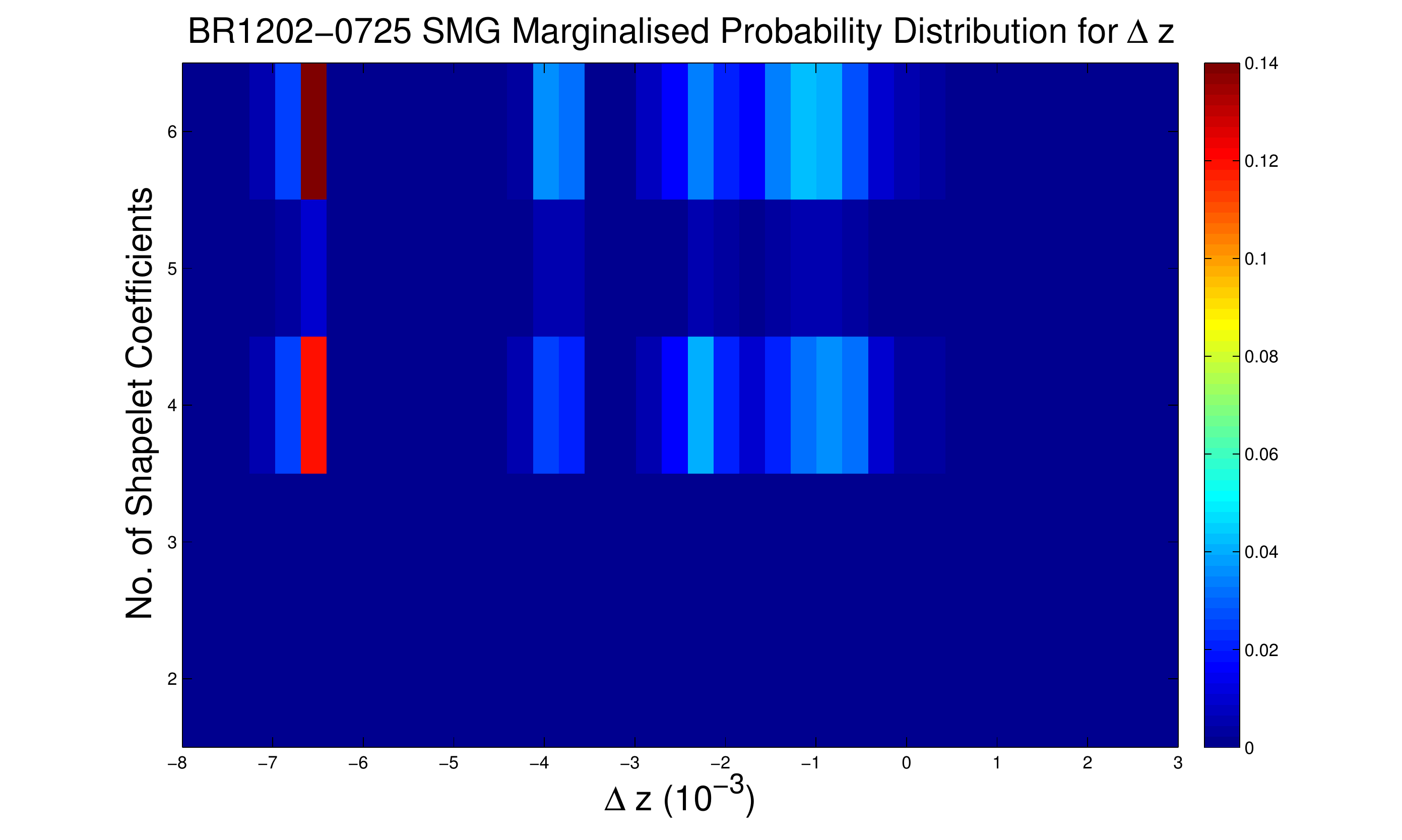} &
\includegraphics[width=80mm]{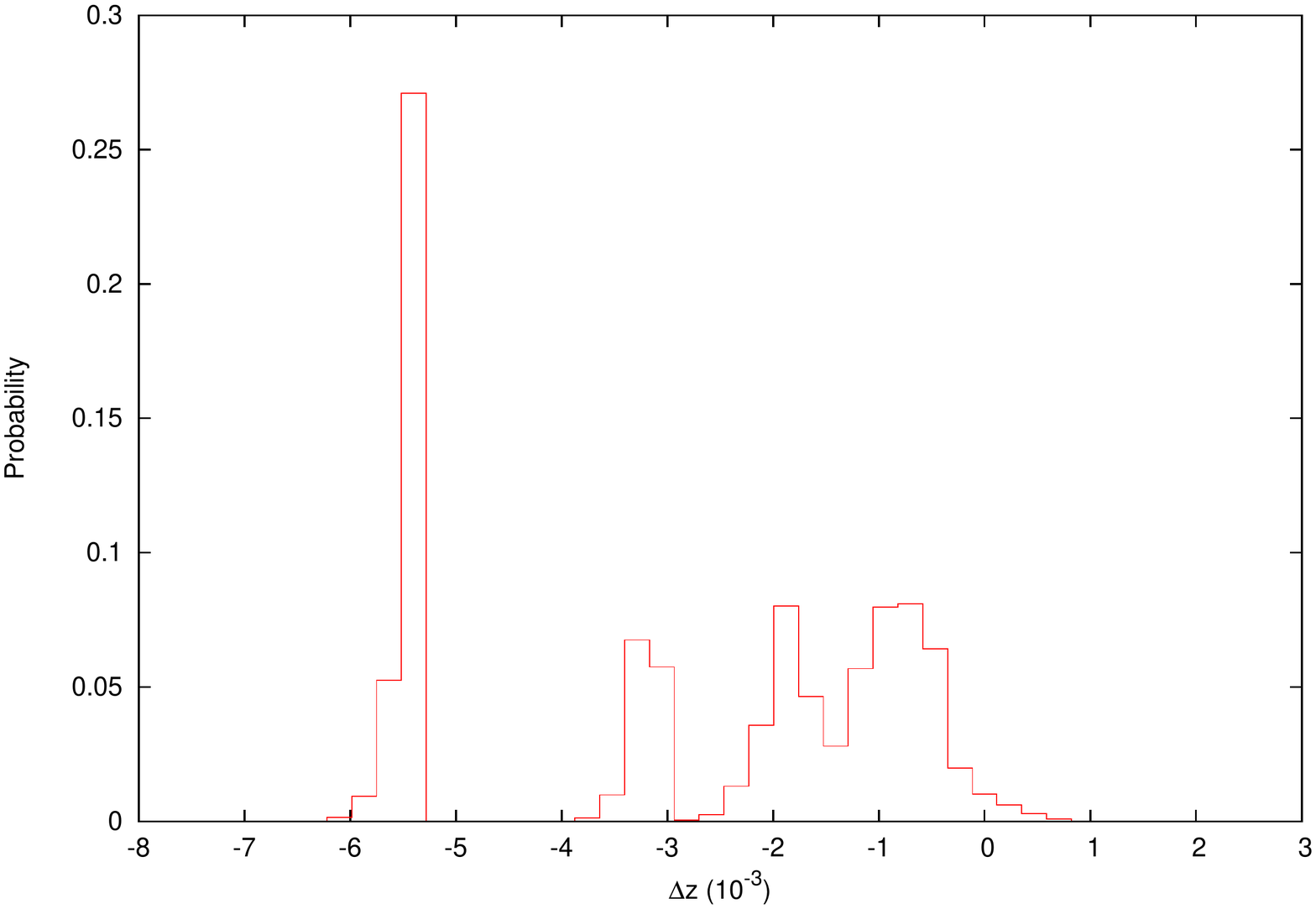}
\end{array}$
\end{center}
\caption{Left: Normalised 2D probability distribution for $n_{\mathrm{max}}$ and $\Delta F/F$ marginalized over shapelet coefficients.   Right: Normalised 1D probability distribution for $\Delta F/F$ marginalized over all $n_{\mathrm{max}}$ and all shapelet coefficients.}
\label{figure:1202marginprofiles}
\end{minipage}
\end{figure*}

The evidence heavily favors two independent line shapes, with a difference in the log evidence of 20, corresponding to a probability of $> 99.99\%$ that the emission lines do not share the same profile.  Due to the non-Gaussian nature of their shape, comparing the line centroids for independent models makes little sense, as the centers are at that stage largely arbitrary, however forcing a single profile upon the pair we find a best fit described by $n_{\mathrm{max}}$ of 3, with FWHM of $700 \pm 26$ km s$^{-1}$.
\begin{table}
\centering
\caption{BR1202$-$0725 SMG Emission Line Parameters from Joint Fit} 
\centering 
\begin{tabular}{c c c} 
\hline\hline 
Line & $\nu_{\mathrm{obs}}$ & $z$ \\[0.5ex] 
	&	GHz			& \\
\hline 
[C~II] &  333.762 $\pm$ 0.009 & 4.69429 $\pm$ 0.00016 \\
CO~($5\to4$) & 101.236 $\pm$ 0.022 & 4.6923 $\pm$ 0.0012 \\
\hline
\end{tabular}
\label{Table:1202SMGjoint} 
\end{table}
Line centroids and redshifts are given in Table \ref{Table:1202SMGjoint}, and correspond to values of $\Delta z = (-2.0 \pm 1.2) \times 10^{-3}$, and once again assuming this change in redshift is the result of only an evolution in the fundamental constants, $\Delta F/F = (-3.5 \pm 2.0) \times 10^{-4}$.
The results of the marginalization process are shown in Fig. \ref{figure:1202marginprofiles}. The posterior for $\Delta z$ displays a large number of distinct peaks resulting from the irregular line shapes producing a large number of likely models across a range of $\Delta z$.  We take the most Gaussian like region to represent our $99\%$ confidence interval for $\Delta z$ of $-1.1\pm1.5$.

\section{Discussion}
\label{section:Discussion}

\subsection{Spatial Distributions of CO and [C~II]}

When calculating the variation in $F$ by comparing the line centroids of different species, one of the most important considerations is whether or not those two species have the same spatial distribution within the host galaxy.
Recent high resolution studies ($\sim$ 50 parsec scale) of star forming regions in M33 \cite{2011A&A...532A.152M} have shown that the CO~(J = $2\to1$) emission line is shifted by $+1.6$ km s$^{-1}$ relative to the [C~II].  These authors find that significant parts of the [C~II] are not traced by the CO because the latter is photo-dissociated in the low-metallicity environment of M33.  \cite{2006A&A...453...77R} compared [C~II] and CO emission on the scale of $\sim$ 300 pc in the spiral arms of M31 and find that whilst overall they trace the same regions, their maxima are not coincident, with the tighter correlation being between the [C~II] and H$_\alpha$, i.e. tracers of star formation. As such, below 300 pc we are not justified in assuming that the two species co-exist spatially.  

Unresolved comparisons of [C~II] and CO (see e.g. \cite{1991ApJ...373..423S}) however have shown good agreement between line profiles on $\sim$ kpc scales.  Our comparison of the [C~II] and CO~(J = $2\to1$) for J1148+5251 showed that there is a 76$\%$ probability that the two emission lines share the same profile, indicating that at least on large scales they may indeed trace out the same spatial volume.  

For the BR1202$-$0725 QSO and SMG however, we found $\sim$65 and $>$ 99.9 $\%$ probabilities that the CO~(J = $5\to4$) and [C~II] emission lines have different intrinsic profiles.  In both cases the CO~(J = $5\to4$) was approximately 30$\%$ wider than the [C~II], which suggests that even on large scales the spatial distribution of these two species is not coherent.  We therefore see that with individual systems showing such discrepancies, any one estimate of $\Delta z$ from a single source will inherently have some unknown error associated with offsets in the distribution of the two species of gas.

Several physical explanations exist to explain such variations; Compact starburst regions can be optically thick in CO, but thin in [C~II], and hence the CO line only samples the outer regions of the galaxy leading to larger line widths for the CO relative to [C~II].  Conversely, the dust emission in such regions can become optically thick in the rest-frame FIR  (at the frequency of the [C~II] line), reducing it's intensity.  Both [C~II] and CO emission has been observed in the photon-dominated regions (PDRs) of molecular clouds \cite{1999RvMP...71..173H}.  Models suggest a separation in the two species, with [C~II] tracing the outer layers of molecular clouds, and CO emission coming from within.  Observations of the warm interstellar medium in the Galaxy \cite{2012A&A...541L..10V} have also show that [C~II] traces a larger region than the CO.

\subsection{The evolution of the fundamental constants}

\begin{table*}
\centering
\caption{Final derived values for $\Delta F/F$} 
\centering 
\begin{tabular}{c c c} 
\hline\hline 
	   			         & $\Delta F/F$ \\
System 				 &     Single Joint Fit   &     Marginalised Result \\
					&	$\pm$ (3$\sigma$ errors) & (99$\%$ Confidence Interval)\\[0.5ex] 
\hline
J1148+5251		  	 & ($ -3.6 \pm 2.4 ) \times 10^{-4}$  & ($ -3.3 \pm 2.3 ) \times 10^{-4}$.  \\
BR1202$-$0725 QSO	 &	($-5 \pm 15) \times 10^{-5}$ & ( $-5 \pm 15) \times 10^{-5}$	\\
BR1202$-$0725 SMG	 &	($-4 \pm 6) \times 10^{-4}$ & $(-4 \pm 6) \times 10^{-3}$\\
\hline 

\hline
\end{tabular}
\label{Table:finalresults} 
\end{table*}

Table ~\ref{Table:finalresults} summarises our results for $\Delta F/F$ for both approaches - calculating the single joint model that maximises the evidence, and marginalising over all possible models.  Errors for the single fit are given as $3\sigma$ values, whilst the result for marginalising over the posterior is given as a 99$\%$ confidence interval.

Our value of $\Delta F/F$ for J1148+5251 of $-3.3 \pm 2.3 \times 10^{-4}$ is the result of a velocity offset with a 99$\%$ confidence interval of $99 \pm 69$ km s$^{-1}$.  Thus in order to account for the offset from zero we would require that either the interstellar medium has an intrinsic velocity dispersion much greater than expected, or the two species are positioned in such a way that their profiles are consistent, but their relative velocities are significantly offset.  
From Fig. \ref{figure:1148profiles}, we can see that the data obtained for the CO~($2\to1$) line only just extends beyond the width of the line.  There is the possibility then that the baseline level for this emission line has an additional component that is not easily visible from the data.  We therefore allowed the continuum component of the [C~II] and CO lines to vary independently for J1148+5251 to see how this affected the outcome of the fit.  The result was that it did not;  The [C~II] and CO fitted continuum components of $+0.43 \pm 0.37$ and $+0.008 \pm 0.009$ respectively, with $\Delta F/F = (-3.3 \pm 0.8) \times 10^{-4}$ as before.

For the BR1202$-$0725 QSO our value for $\Delta F/F$ of ( $-5 \pm 15) \times 10^{-5}$ results from a velocity offset of $15 \pm 45 km s^{-1}$, which is consistent with our estimates of the intrinsic velocity dispersion of the two species.

Finally for the BR1202$-$0725 SMG the best fit model we found a velocity offset of $-120 \pm 60$ km s$^{-1}$, whilst the marginalised posterior had multiple peaks between a range of $-375$ and $+30$ km s$^{-1}$.  This is the result of attempting to fit a joint model to the two emission lines where the evidence strongly favours different profiles for each.  We note that the range of the offset in velocity seen in the marginalised posterior is roughly consistent with the difference in the widths of their profiles when fitted independently ($\sim$ 300 km s$^{-1}$)

If we want to reach a limit of $\Delta F/F < 10^{-6}$ using CO and [C~II] we will therefore need to sample large numbers of high $z$ objects in order to reduce the error associated with the random motions of the gas to $\sim$ 0.3 km s$^{-1}$ and address other factors such as differing (both gas and dust) opacities.  To go from even a modest estimate of this random motion of $10$ km s$^{-1}$, this would require $\sim$ 1000 sources, assuming Gaussian statistics.  Alternatively,  spatially resolved imaging could be performed on a number of sources, thereby providing multiple, independent line-of-sight measurements for each source.   We could also choose to use two species whose spatial distributions are believed to be more closely tied, but that are still detectable with high signal to noise at comparable look back times, for example \citet{2012ApJ...753..102W} use a combination of C~I and CO~($7\to6$) emission from a $z=2.79$ QSO host galaxy to derive limits on $\Delta F/F$ of $6.9 \pm 3.7 \times 10^{-6}$.  

Finally if we compare our results to the proposed angular dipole distribution \cite{2011PhRvL.107s1101W,2012MNRAS.422.3370K}, J1148+251 lies in a region with $\cos\theta \sim -0.9$ with $\theta$ the angular separation with respect to the dipole, and so we would expect $\Delta \alpha/\alpha \sim -0.9\times10^{-5}$.  Given $\Delta F/F = 2\Delta \alpha/\alpha - \Delta \mu/\mu$, if we take $ \Delta \mu/\mu=0$ we find $\Delta \alpha/\alpha =  (-1.8 \pm 1.2 ) \times 10^{-4}$ which is consistent with the value predicted by the dipole.  BR1202$-$0725 however lies almost orthogonal to the dipole with $\cos\theta \sim 0.2$ and so does not produce any stringent tests on the claim.

\subsection{The origin of $\Delta z$}

Out of the three systems investigated only the [C~II] and CO(J = 2$\to$1) transition lines from J1148+5251 were seen to be consistent with one another within the limits of the data, however we note that the CO(J = 2$\to$1) emission line also had the lowest signal to noise of any of the data presented here.  The CO~(J = $5\to4$) and [C~II] emission lines from the BR1202$-$0725 QSO and SMG however were both seen to have intrinsically different profiles, where in both cases the CO~(J = $5\to4$) was approximately 30$\%$ wider than the [C~II].
As such we propose that for individual systems the differences in the redshifts $\Delta z$ of the CO and [C~II] emission lines cannot be viewed as being solely due to the evolution of the fundamental constants, but must include a contribution owing to the intrinsically different spatial distributions of the two species of gas within the galaxy.

\section{Conclusions}
\label{section:Conclusions}

We have presented a series of sensitive observations of molecular CO and [C~II] emission in 3 high redshift galaxies; the QSO host J1148+5251 at $z=6.4$ and the QSO host BR1202$-$0725 at $z=4.7$ along with it's companion SMG.  We have used these to quantify the differences in the two emission line profiles for each system and thus put constraints on the quantity $\Delta z = z_{CO} - z_{C_{II}}$ in two distinct ways; Calculating the Bayesian evidence for a joint model using shapelets and determining the separation in redshift between those models, and by marginalising over all model parameters to calculate the posterior distribution associated with $\Delta z$ independent of any particular model choice.  

We found that the line profiles for the CO~($5\to4$) transition for the BR1202$-$0725 QSO and SMG were inconsistent with the [C~II], being approximately 30$\%$ wider whereas the  CO~($2\to1$) and [C~II] emission lines for J1148+5251 were consistent within the limits of the data.  This suggests that, in agreement with existing high resolution observations, whilst the CO and [C~II] might largely trace the same components of the host galaxy, there are still observable differences between their line profiles.  As such a direct conversion from $\Delta z$ to the quantity $\Delta F/F$ in order to constrain the evolution of fundamental constants for an individual system using these two species of gas is problematic, with our results being fundamentally limited by this unknown factor that represents the relative distributions and velocities of the gas within the host galaxy.  Such issues are potentially eliminated either with large sample sizes to reduce the random error, or by observing different species that are more closely linked.  In either case, the sensitivity of ALMA will allow large numbers of CO and [C~II] detections at high redshift to mitigate the issue, allowing us to reduce the uncertainty in the evolution of the fundamental constants at these distant look back times to the level of $\sim 10^{-6}$, on a par with methods used at lower redshifts.

\end{document}